 \documentclass[final,5p,times,twocolumn]{elsarticle}

\usepackage{amssymb}

\usepackage{multirow}
\usepackage[utf8]{inputenc} 
\usepackage[T1]{fontenc}
\usepackage{graphicx}
\usepackage{float}
\usepackage[dvipsnames]{xcolor}
\usepackage{caption}
\usepackage{subcaption}
\usepackage{hyperref}

\hypersetup{
  colorlinks   = true,    
  urlcolor     = blue,    
  linkcolor    = blue,    
  citecolor    = red      
}
\input glyphtounicode
\journal{Journal Name}

\begin{document}

\begin{frontmatter}



\title{
DILATION TWO EMBEDDING ONE-BY-ONE PARTICULAR SUB-QUADTREE INTO $\mathbf{M}$-DIMENTIONAL CROSSED CUBES 
}


\author[mymainaddress]{Aymen Takie Eddine Selmi}
\ead{aymen.selmi@univ-biskra.dz}
\author[mysecondaryaddress]{Mohamed Faouzi Zerarka}
\ead{faouzi.zerarka@univ-biskra.dz}

\author[mysecondaryeaddress]{Abdelhakim Cheriet}
\ead{ahcheriet@gmail.com}

\address[mymainaddress]{Departement of computer science and LESIA Laboratory, Mohamed Khider University of Biskra, Algeria}
\address[mysecondaryaddress]{Departement of computer science, Mohamed Khider University of Biskra, Algeria.}
\address[mysecondaryeaddress]{Department of computer science and information technology, Kasdi Merbah University of Ouargla, Algeria.}

\begin{abstract}
In the parallel processing field, graph embedding is motivated by simulation interconnection networks to another. The quadtree is an important technique used to present spatial data and is used in many application domains, especially computer vision and image processing. Researchers are interested in the construction and manipulation of quadtrees on parallel machines. The crossed cubes consider an alternative to the ordinary hypercube. It offers many attractive properties. Significantly, it reduces diameter by a factor of 2 that of the ordinary cubes. Moreover, the crossed cubes have a great capacity to simulate other architectures. This paper is interested in the one-by-one dilation two embedding of a particular sub-quadtree graph into m-dimensional crossed cubes.\\

\end{abstract}


\end{frontmatter}


\section{Introduction}

For a machine based on parallel architecture, the choice of a good topology is linked to a set of attractive and popular properties such as degree, diameter, connectivity, regularity, embeddability, and fault tolerance.

The quadtree is an important technique used in many application domains such as geographic information systems, image processing, computer graphics, robotics \cite{14,15,17,29,30}. The quadtree is a simple topology; it is a connected graph with a cycle in which each internal vertex has four children. Suppose all left child breaks down; quadtree will be a particular sub-quadtree. This technique has a set of limits. It has a bisection width equal to 1 \cite{23}. Moreover, its connectivity equal to 1 \cite{23}. Therefore, many researchers studied the embedding of quadtrees into another interconnection network \cite{31,32,33,9}.

Parallel machines based on hypercube topology offer an interconnection topology with attractive properties: symmetry, logarithmic diameter, fixed degree, high connectivity, Hamiltonian, fault tolerance, extensibility, and embeddability of other topologies \cite{5,6,7,8,9,24,34}. Several researchers propose several versions of the hypercube to improve its capacity. In the papers by Ahmed El-Amawy\cite{2} and Preparata \& Vuillemin \cite{3}, an attractive version of the hypercube called the crossed cubes is proposed by Efe \cite{4}. This version pays attention because of the similarity in properties with the ordinary hypercube. Also, it offers many popular attractive properties over the ordinary hypercube \cite{9,10,11,12,13,22,35}. Especially, it reduces diameter by a factor of 2 \cite{4}. Moreover, the crossed cubes have a great capacity to simulate other architectures \cite{9,24,25,26,27,28}.
The problem of simulation of one interconnection network to another is essential in the field of parallel computing. The embedding capabilities are important in evaluating an interconnection network. 

Let \emph{G} and \emph{H} be two graphs such that an embedding of \emph{G} into \emph{H} is a pair (\emph{f}, \emph{R}) where \emph{f} is an injective mapping vertex \emph{V}(\emph{G}) into vertex \emph{V}(\emph{H}). \emph{R} is an injective mapping associating with each edge [\emph{u}, \emph{v}] from $G$ at a path \emph{R}(\emph{u}, \emph{v}) which connects \emph{f}(\emph{u}) and  \emph{f}(\emph{v}) \cite{1,18,20}.

This paper aims to construct the dilation two embedding one-by-one particular sub-quadtree into $m$-dimensional crossed cubes. This paper is organized as follows; first, we introduce a few preliminary definitions of the particular sub-quadtree graph and the crossed cubes graph. Section 3 presents the construction of dilation two one-by-one particular sub-quadtree into $m$-dimensional crossed cubes; then, we offer the validation of our new function. Finally, section 4 concludes the paper and discusses some possible future work.
\section{PRELIMINARIES}

\subsection{Definition 01}

The quadtree is an undirected graph \emph{QT}$_{n}$ of $(4^{n} - 1)/3$ vertices. Every vertex of depth less than $n$ has four children, and every vertex of depth equal to $n$ is a leaf. We assume each left child breaks down, quadtree reduced to a particular sub-quadtree graph denoted \emph{PQT$_{n}$}. Therefore, quadtree will become an undirected graph \emph{PQT$_{n}$} of $(3{^n} - 1)/2$ vertices. Let $p$ be a positive integer, each vertex in \emph{PQT$_{n}$} is a string of length $p$ denoted \emph{AV} $=$ \emph{Aa$_{p-1}$suff$_{i}$}; where \emph{A}$_{p-1}$ $=$  \emph{a$_{1}$a$_{2}$$\dots$a$_{j}$$\dots$a$_{p-1}$}, a$_{j}= \overline{1, 3}$ and \emph{suff$_{i}$} $=$ $i$  / $i = \overline{1, 3}$. The root can represent by address $a_{1}$ $=$ $0$. We represent an edge between a vertex parent and one of its children as follows, \emph{A$_{p-1}$}-\emph{A$_{p-1}$suff$_{i}$}.

\subsection{Definition 02 \cite{4}} \label{sssec:num1}
Both the $m$-dimensional hypercube denoted by \emph{Q$_{m}$} and the crossed cubes \emph{CQ$_{m}$} are undirected graphs consisting of the same set of vertices. A binary string of length $m$ labels each vertex in \emph{Q$_{m}$}(\emph{CQ$_{m}$}). In \emph{Q$_{m}$}, two vertices are adjacent if and only if the binary representation of their labels differs in exactly one-bit position. While in the crossed cubes, two binary strings \emph{x} $=$ \emph{x$_{1}$x$_{0}$}, \emph{y} $=$ \emph{y$_{1}$y$_{0}$ } of length two are pair-related if and only if $(\emph{x}, \emph{y}) \in \lbrace (00, 00), (10, 10), (01, 11), (11, 01) \rbrace $.
The $m$-dimensional crossed cubes \emph{CQ$_{m}$} is defined recursively:\

\begin{itemize}
\item \emph{CQ$_{1}$} is the complete graph on two vertices with labels 0 and 1.
\item If $m$ $>$ 1: \emph{CQ}$_{m}$ consists of two sub-cubes, 0\emph{CQ}$_{m-1}$ and 1\emph{CQ$_{m-1}$}. Two vertices $u$, $v$ such that $u$ $=$ $0u_{m-2}...u_{0} \in$ 0\emph{CQ}$_{m-1}$, and $v$ $=$ $1v_{m-2}...v_{0} \in$ 1\emph{CQ}$_{m-1}$ are adjacent, if and only if:\\
\emph{u}$_{m-2}$ $=$ \emph{v}$_{m-2}$ if $m$ is even\\
\emph{u}$_{2i+1}$\emph{u}$_{2i}$, \emph{v}$_{2i+1}$\emph{v}$_{2i}$ are pair-related.

\end{itemize}

\subsection{Definition 03 \cite{20,21}}
Let \emph{G} and \emph{H} be two simple undirected graphs. An embedding of the graph \emph{G} into graph \emph{H} is an injective mapping \emph{f} from the vertices of \emph{G} to the vertices of \emph{H}. Four cost functions, dilation, congestion, expansion, and load factor, often measure the quality of an embedding. In this paper, we interest in dilation. The dilation of the embedding is the maximum distance between \emph{f}(\emph{y}) and \emph{f}(\emph{z}) taken over all edges (\emph{y}, \emph{z}) of \emph{G}.

\subsection{Notations}
A particular sub-quadtree \emph{PQT}$_{n}$  is produced by three copies of \emph{PQT}$_{n-1}$ prefixed respectively by $01$\emph{PQT}$_{n-1}$, $02$\emph{PQT}$_{n-1}$, $03$\emph{PQT}$_{n-1}$, and a root prefixed by $0$\emph{PQT}$_n$.

A crossed cubes \emph{CQ}$_{m}$ $=$ (\emph{O}, \emph{E}), with \emph{E} set of vertices and \emph{O} set of edges.

Let $B \in E$ such that: $B$ $=$ \textcolor{ForestGreen}{$C$}\textcolor{red}{\emph{pref}$_{j}$}$X_{3}X_{2}X_{1}X_{0}$; $\textcolor{ForestGreen}{C} = {\textcolor{ForestGreen}{b_{r-1}}, \textcolor{ForestGreen}{\bar{b}_{r-1}}, \phi}$; $adrr = X, Y, Z$, $adrr$ $= X_{1}X_{0}$ of length two and ($X, Y$) or ($Y, Z$) are pair-related; \textcolor{red}{\emph{Pref}$_{j}$} $=$ \textcolor{red}{$b_{k}\dots b_{m-4}$} / $j$ $=$ $\overline{0, 3}$ respectively if $b_{r}b_{r+1}, b_{r}\bar{b}_{r+1}$, $\bar{b}_{r}b_{r+1}$, $\bar{b}_{r}\bar{b}_{r+1}$. The number of super nodes \emph{CQ}$_{4}$ is equal to 2$^{m-4}$.

\emph{CQ}$_{m}$ is produced as follows:\\
Where $C = \phi$: \emph{CQ}$_{m}$ is produced by four copies of \emph{CQ}$_{m-2}$  prefixed respectively by $00$\emph{CQ}$_{l-2}$, $01$\emph{CQ}$_{l-2}$, $10$\emph{CQ}$_{l-2}$, $11$\emph{CQ}$_{l-2}$.\\
Where $C \neq \phi$: \emph{CQ}$_{m}$ is produced by two copies of \emph{CQ}$_{m-1}$  prefixed respectively by $0$\emph{CQ}$_{m-1}$, $1$\emph{CQ}$_{m-1}$ in other word: $00$\emph{CQ}$_{m-2}$, $01$\emph{CQ}$_{m-2}$, $10$\emph{CQ}$_{m-2}$, $11$\emph{CQ}$_{m-2}$.
\section{DILATION TWO EMBEDDING ONE-BY-ONE PARTICULAR SUB-QUADTREE INTO M-DIMENSIONAL CROSSED CUBES}

 This section describes our new function, which allows the embedding one-by-one particular
 sub-quadtree \emph{PQT$_{n}$} into $m$-dimensional crossed cubes \emph{CQ}$_{m}$. We can resume this function as follows:
\begin{itemize}
\item Determine the dimension of the crossed cubes \emph{CQ}$_{m}$.
\item One-by-one vertex embedding of \emph{PQT$_{n}$} into \emph{CQ}$_{m}$. 
\item Dilation two embedding one-by-one all edges of \emph{PQT$_{n}$} onto paths in \emph{CQ}$_{m}$.
\end{itemize}

\subsection{Dimension of \emph{CQ}$_{m}$} 

The dimension of the crossed cubes $m$ related by $n$ the height of a particular sub-quadtree in which:
\begin{itemize}
\item Where $n \leq 8$: $m = log_{2}((3^{n} - 1) / 2)) / log_{2}(2) $
 \item Where $n > 8$: $m = (n-8)*2 + 12$
\end{itemize}
\subsection{One-by-one vertex embedding}
\begin{figure}[H]
\centering
\includegraphics[width=8cm,height=8cm,keepaspectratio]{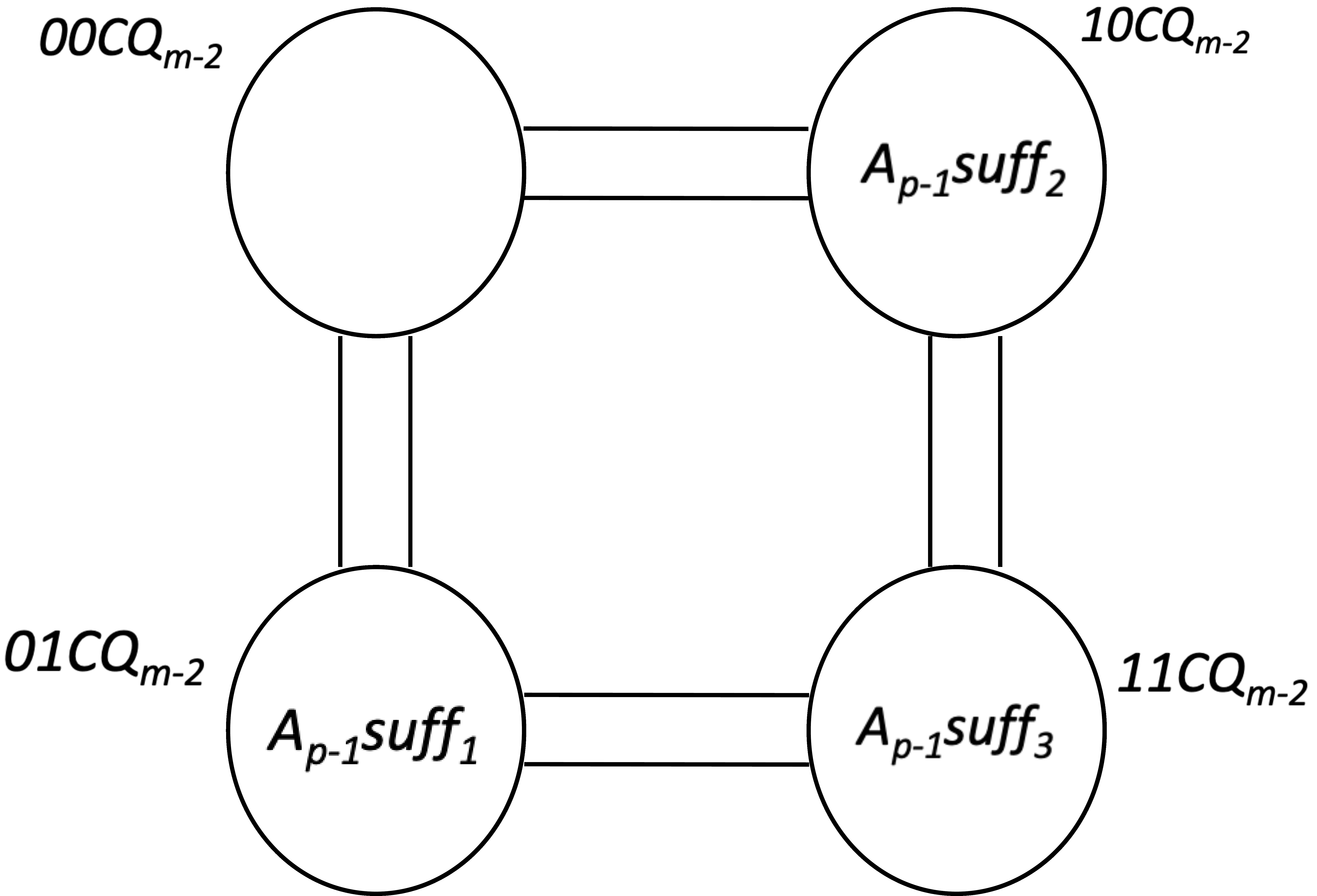}
\caption{Vertex embedding Situation 1} \label{fig:ns1}
\end{figure}

\begin{figure}[H]
\centering
\includegraphics[width=8.5cm,height=9cm,keepaspectratio]{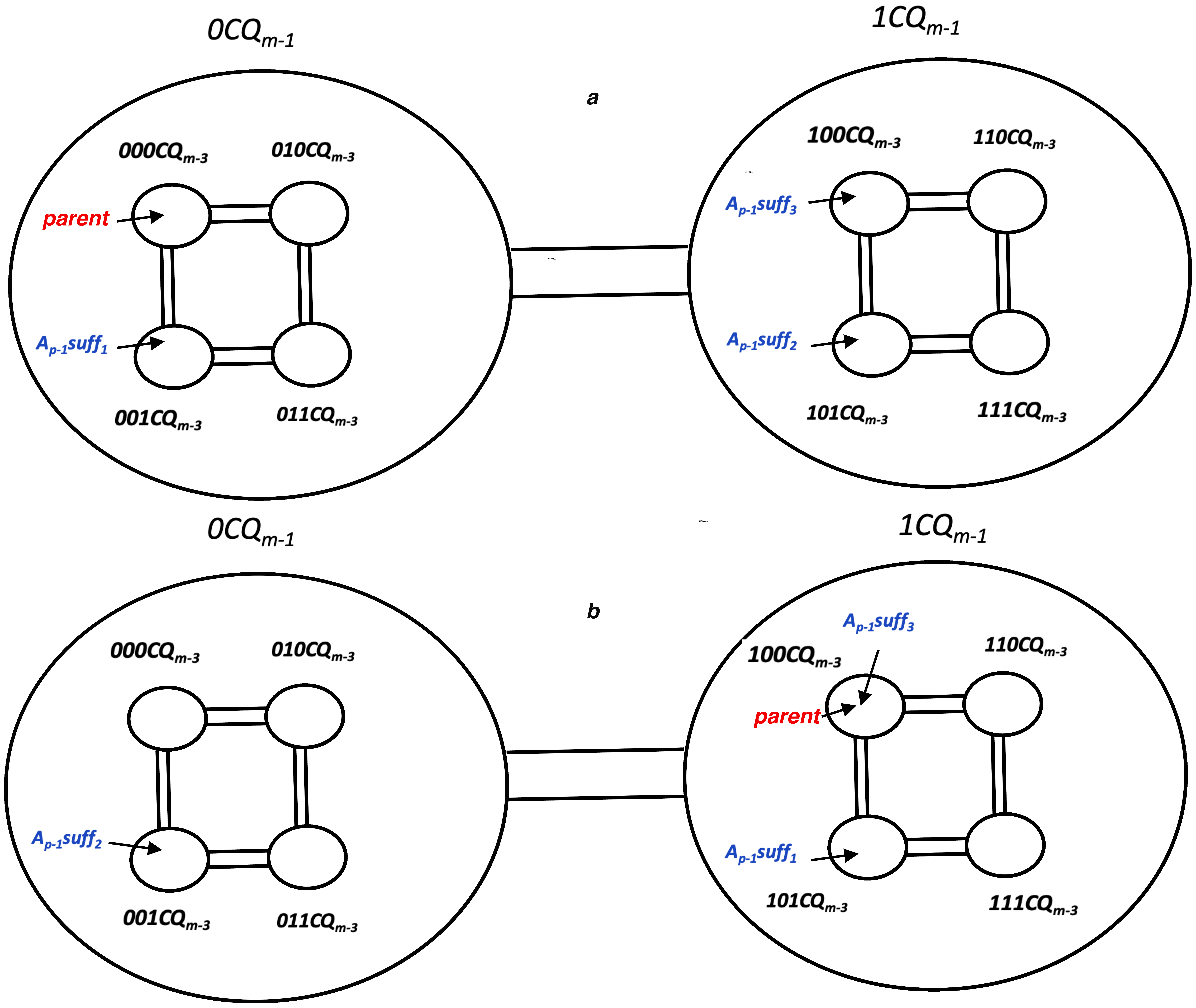}
\caption{Vertex embedding situation 2, case 1.} \label{fig:ns2c1}
\end{figure}

\begin{figure}[H]
\centering
\includegraphics[width=8.5cm,height=9cm,keepaspectratio]{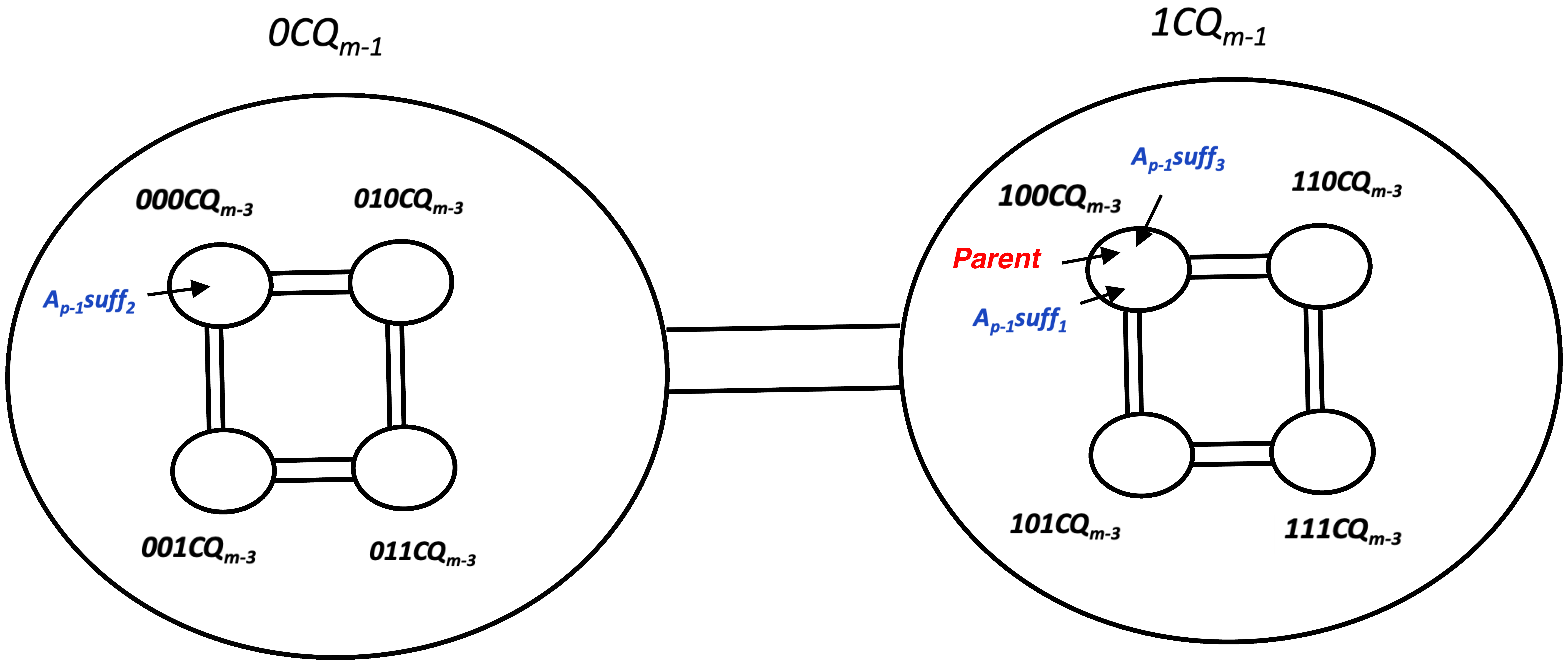}
\caption{Vertex embedding situation 2, case 2.}\label{fig:ns2c2}
\end{figure}

\begin{figure}[H]
\centering
\includegraphics[width=8.5cm,height=9cm,keepaspectratio]{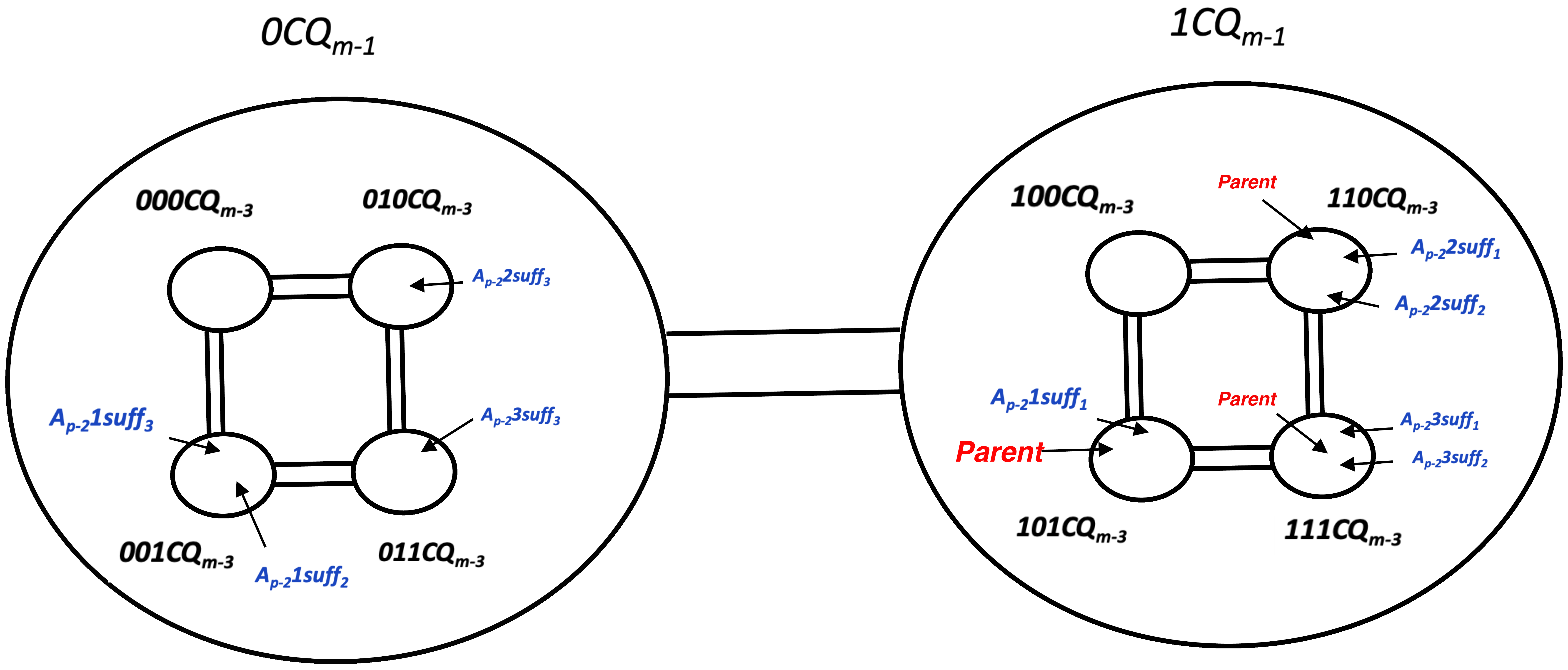}
\caption{Vertex embedding situation 2, case 3.} \label{fig:ns2c3}
\end{figure}

The one-by-one vertex embedding of \emph{PQT$_{n}$} into \emph{CQ}$_{m}$ is done in the following way:\\
For $n=3$: The basic function \emph{f} of this one by one vertex embedding is produced as follows:
\begin{itemize}
\item 	\emph{Prem(0) := pref$_{0}$$00$}
\item 	\emph{f(A$_{p-1}$suff$_{1}$) := Pref$_{1}$$X$}
\item 	\emph{f(A$_{p-1}$suff$_{2}$) := Pref$_{2}$$Y$}
\item   \emph{f(A$_{p-1}$suff$_{3}$) := Pref$_{3}$$Z$}
\end{itemize}
For $n>3$: the one-by-one vertex embedding is done in two situations. The first one is when $C=\phi$, we use the basic function \emph{f} (figure \ref{fig:ns1}).
The second is when $C\neq \phi $; a function \emph{f}$_{1}$ of this one-by-one vertex embedding. Thus, there three cases in this situation; the following rules of case 1 shown in figure \ref{fig:ns2c1} produce \emph{f}$_{1}$:
\begin{itemize}
\item 	\emph{f$_{1}$(A$_{p-1}$suff$_{1}$) := $0$Pref$_{1}$$00X$}
\item 	\emph{f$_{1}$(A$_{p-1}$suff$_{2}$) := $1$Pref$_{1}$$00Y$}
\item   \emph{f$_{1}$(A$_{p-1}$suff$_{3}$) := $1$Pref$_{0}$$00Z$}
\end{itemize}
\hspace{2,2cm} OR
\begin{itemize}
\item 	\emph{f$_{1}$(A$_{p-1}$suff$_{1}$) := $\bar{0}$Pref$_{1}$$00X$}
\item 	\emph{f$_{1}$(A$_{p-1}$suff$_{2}$) := $\bar{1}$Pref$_{1}$$00Y$} 
\item   \emph{f$_{1}$(A$_{p-1}$suff$_{3}$) := $1$Pref$_{0}$$00Z$}
\end{itemize}
The following rules of case 2 shown in figure \ref{fig:ns2c2} produce \emph{f}$_{1}$:
\begin{itemize}
\item \emph{f$_{1}$(A$_{p-1}$suff$_{1}$) := $\bar{0}$Pref$_{0}$$00X$}
\item \emph{f$_{1}$(A$_{p-1}$suff$_{2}$) := $\bar{1}$Pref$_{0}$$00Y$}
\item \emph{f$_{1}$(A$_{p-1}$suff$_{3}$) := $1$Pref$_{0}$$00Z$}
\end{itemize}
The following rules of case 3 shown in figure \ref{fig:ns2c3} produce \emph{f}$_{1}$ ($t$ $=$ 2, 3):
\begin{itemize}
\item 	\emph{f$_{1}$(A$_{p-1}$suff$_{1}$) := $\bar{0}$Pref$_{1}$$00X$}
\item 	\emph{f$_{1}$(A$_{p-1}$suff$_{2}$) := $\bar{1}$Pref$_{1}$$00Y$}
\item   \emph{f$_{1}$(A$_{p-1}$suff$_{3}$) := $\bar{1}$Pref$_{1}$$00Z$} 
\end{itemize}
\hspace{2,2cm} OR
\begin{itemize}
\item 	\emph{f$_{1}$(A$_{p-1}$suff$_{1}$) := $\bar{0}$Pref$_{t}$$00X$}
\item 	\emph{f$_{1}$(A$_{p-1}$suff$_{2}$) := $1$Pref$_{t}$$00Y$} 
\item   \emph{f$_{1}$(A$_{p-1}$suff$_{3}$) := $\bar{1}$Pref$_{t}$$00Z$}
\end{itemize}

\begin{figure}[H]
\centering
\includegraphics[width=6cm,height=6cm,keepaspectratio]{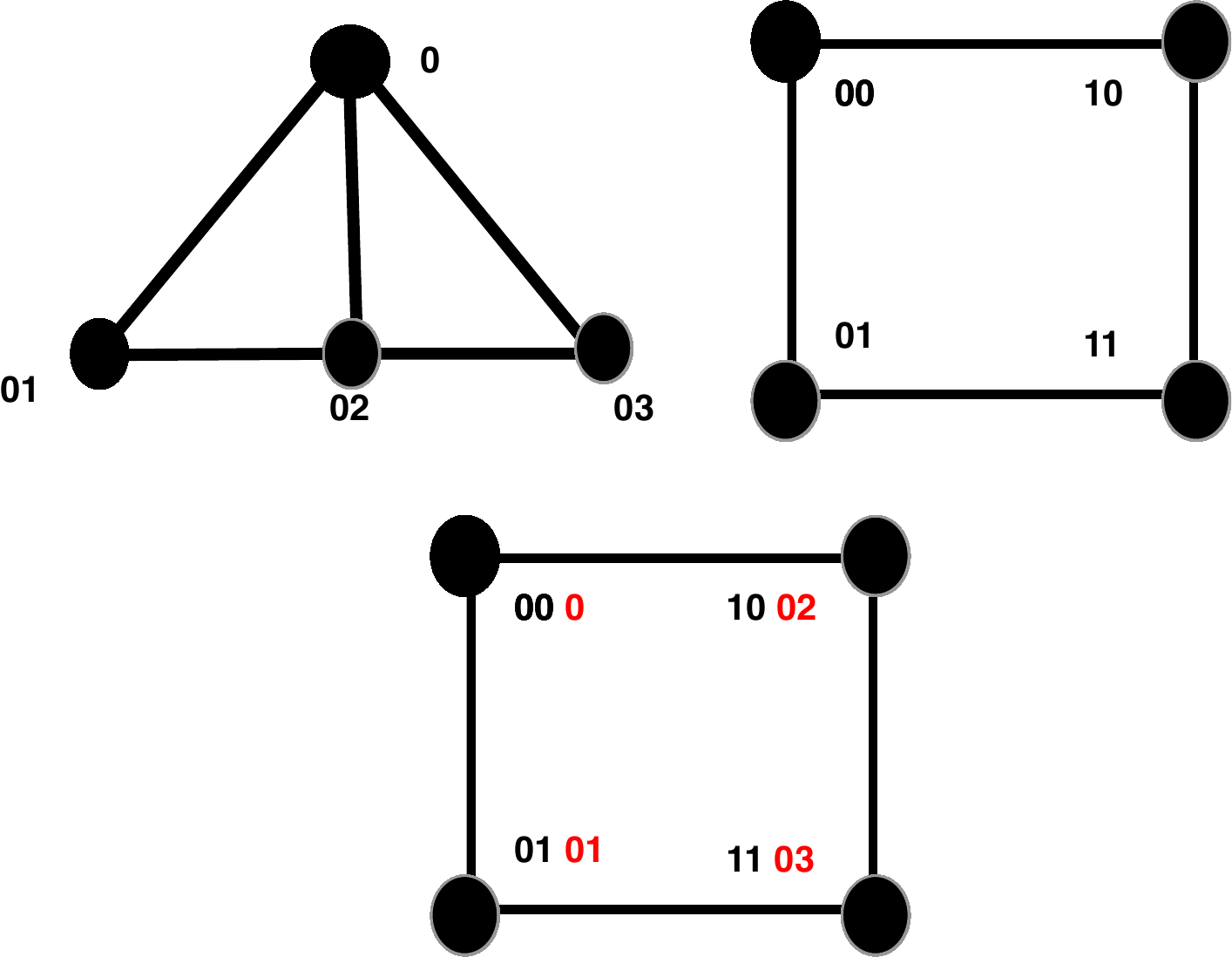}
\caption{Nodes embedding graph of \emph{PQT}$_{2}$ into \emph{CQ}$_{2}$.}
\label{fig:n2}
\end{figure}

\begin{figure}[H]
\centering
\includegraphics[width=8cm,height=8cm,keepaspectratio]{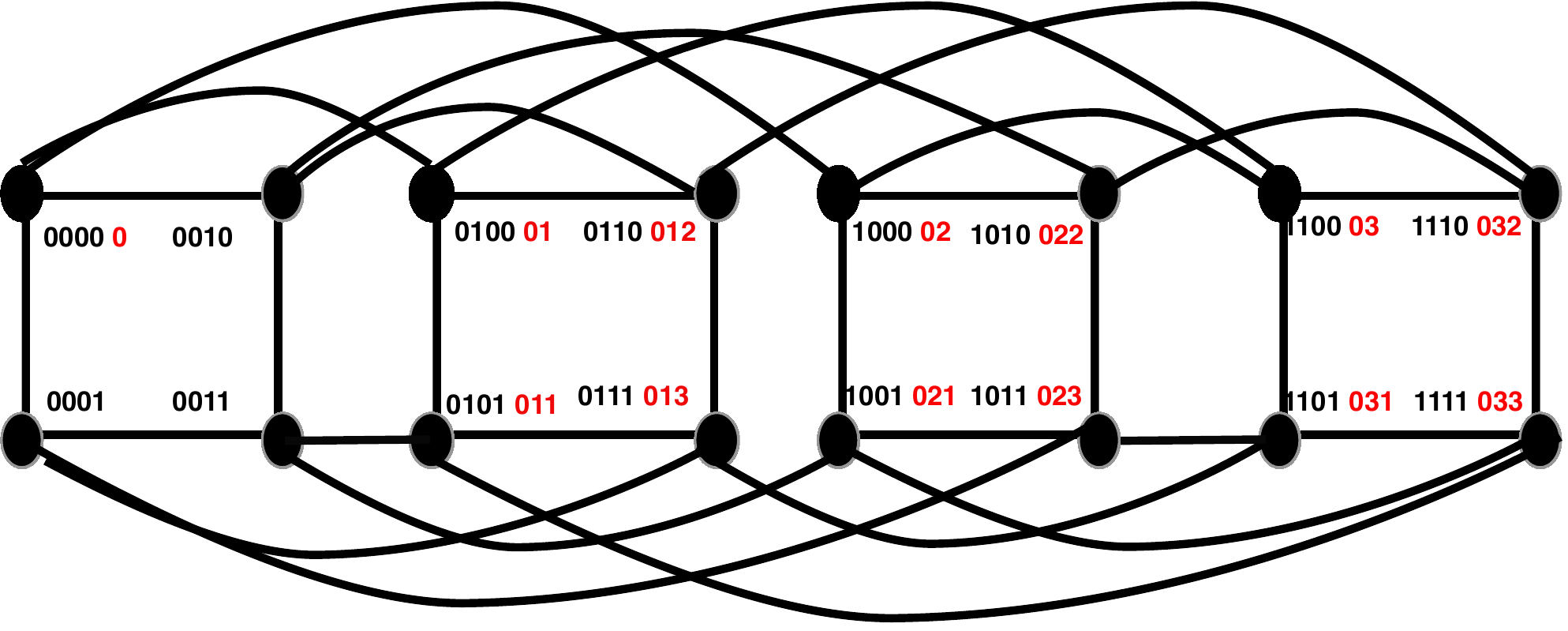}
\caption{Nodes embedding graph of \emph{PQT}$_{3}$ into \emph{CQ}$_{4}$.}
\label{fig:n3}
\end{figure}

\begin{figure}[H]
\centering
\includegraphics[width=8cm,height=8cm,keepaspectratio]{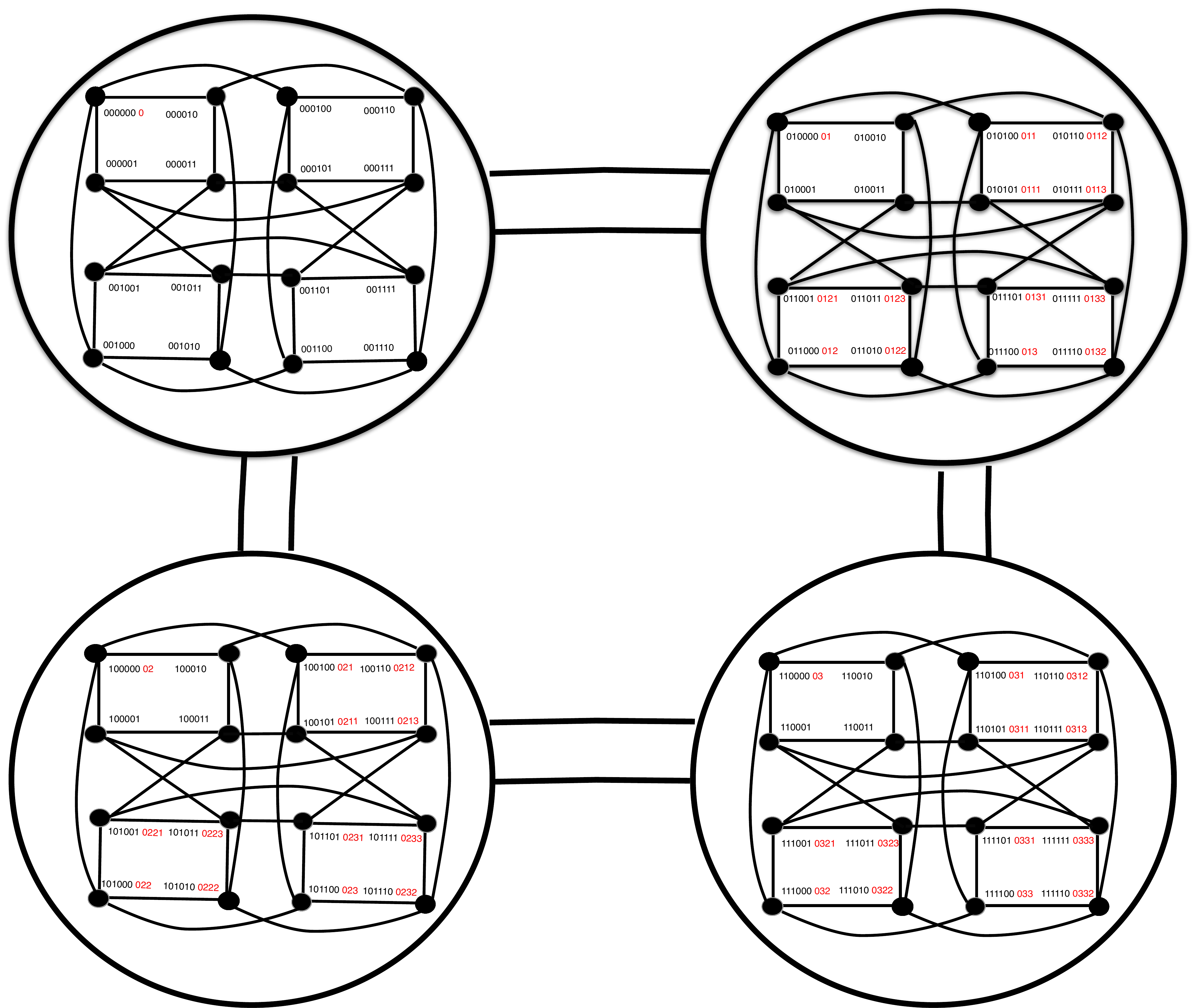}
\caption{Nodes embedding graph of \emph{PQT}$_{4}$ into \emph{CQ}$_{6}$.}
\label{fig:n4}
\end{figure}

\textbf{Lemma 1.}
For $n<5$, a particular sub-quadtree \emph{PQT$_{n}$} is one-by-one vertex embedding into $m$-dimensional crossed cubes \emph{CQ}$_{m}$.\\
 \textbf{Proof.}
 We prove lemma 1 by induction on $n$.\\
\textbf{Base.}
For $n = 2$: as shown in figure \ref{fig:n2}.\\
For $ n = 3, 4$: level’s 1, 2 nodes of \emph{PQT}$_{3}$, \emph{PQT}$_{4}$ are respectively embedded into \emph{CQ}$_{4}$, \emph{CQ}$_{6}$ using the rules specified in table \ref{t1}, table \ref{t2}, shown in figure \ref{fig:n3}, figure \ref{fig:n4}.\\
\begin{table}[H]
\centering
\resizebox{\columnwidth}{!}{%
\begin{tabular}{|c|c|c|c|}
\hline
\emph{Root}   & \emph{Prem(root)} & $0$\emph{suff}$_{1}$ & $01$\emph{CQ}$_2$ \\ \hline
$\textcolor{blue}{0}$      & $\textcolor{red}{00}00$       & $0\textcolor{blue}{1}$     & $\textcolor{red}{01}00$  \\ \hline
$0$\emph{suff}$_{2}$& $10$\emph{CQ}$_2$      & $0$\emph{suff}$_{3}$ & $11$\emph{CQ}$_2$ \\ \hline
$0\textcolor{blue}{2}$     & $\textcolor{red}{10}00$       & $0\textcolor{blue}{3}$     & $\textcolor{red}{11}00$  \\ \hline
\end{tabular}
}
\caption{Level’s 1, 2 nodes embedding of \emph{PQT}$_{3}$ into \emph{CQ}$_{4}$}
\label{t1}
\end{table}

\begin{table}[H]
\centering
\resizebox{\columnwidth}{!}{%
\begin{tabular}{ |c|c|c|c|}
\hline
\emph{Root}   & \emph{Prem(root)} & $0$\emph{suff}$_{1}$ & $01$\emph{CQ}$_4$ \\ \hline
$\textcolor{blue}{0}$      & $\textcolor{red}{00}0000$       & $0\textcolor{blue}{1}$     & $\textcolor{red}{01}0000$  \\ \hline
$0$\emph{suff}$_{2}$& $10$\emph{CQ}$_4$      & $0$\emph{suff}$_{3}$ & $11$\emph{CQ}$_4$ \\ \hline
$0\textcolor{blue}{2}$     & $\textcolor{red}{10}0000$       & $0\textcolor{blue}{3}$     & $\textcolor{red}{11}0000$  \\ \hline
\end{tabular}
}
\caption{Level’s 1, 2 nodes embedding of \emph{PQT}$_{4}$ into \emph{CQ}$_{6}$.}
\label{t2}
\end{table}

\textbf{Induction hypothesis}

Suppose that for $k \leq n - 1$, \emph{PQT}$_{k}$ is one-by-one vertex embedding into \emph{CQ}$_{l}$ with $l < m$  is true.\\
Let us now prove that it is true for $k = n$.

The root $0$ is embedded into $00$\emph{CQ}$_{l-2}$ by using \emph{Prem(root)}. Nodes \emph{A}$_{k-1}$\emph{suff}$_{1}$, \emph{A}$_{k-1}$\emph{suff}$_{2}$, \emph{A}$_{k-1}$\emph{suff}$_{3}$  are respectively embedded into $01$\emph{CQ}$_{l-2}$, $10$\emph{CQ}$_{l-2}$, $11$\emph{CQ}$_{l-2}$ using the basic function \emph{f} (induction hypothesis, shown in figure \ref{fig:ns1}).\\

\begin{figure*}[ht]
\centering
\includegraphics[width=18cm,height=19cm,keepaspectratio]{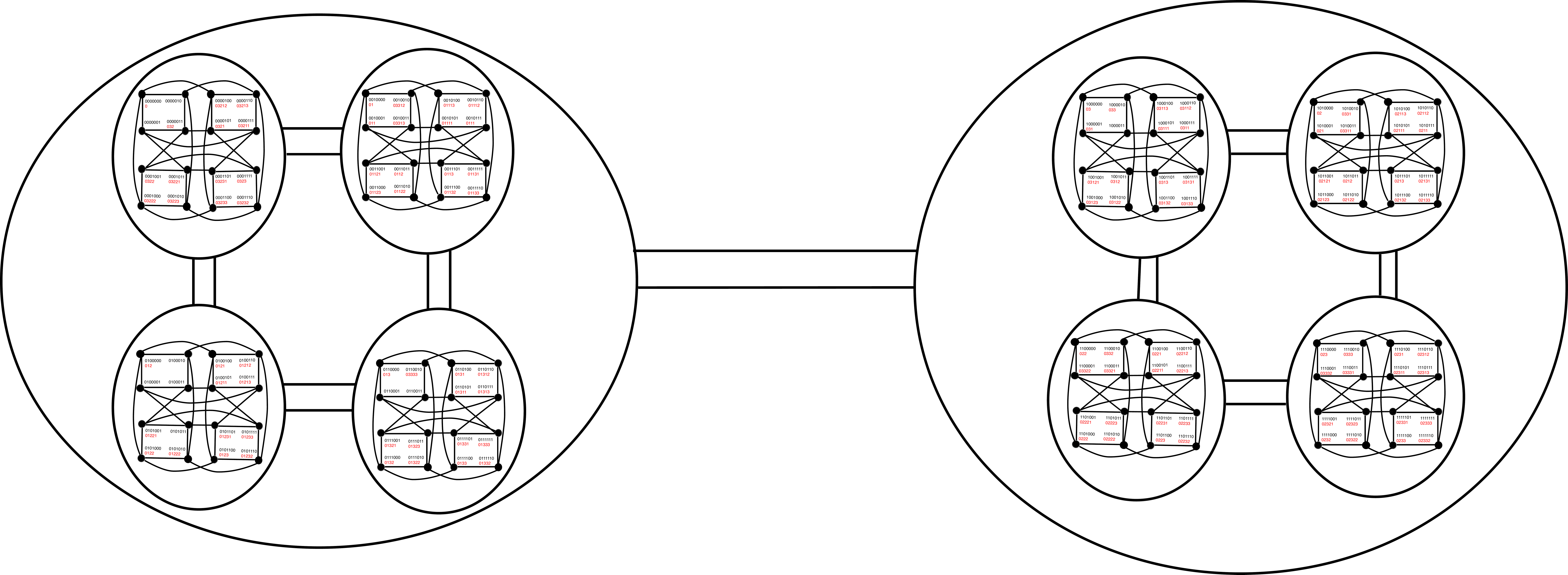}
\caption{Nodes embedding graph of \emph{PQT}$_{5}$ into \emph{CQ}$_{7}$.}
\label{fig:n5}
\end{figure*}

\textbf{Lemma 2.} For $n=5$, a particular sub-quadtree \emph{PQT$_{n}$} is one-by-one vertex embedding into $m$-dimensional crossed cubes \emph{CQ}$_{m}$.\\
\textbf{Proof.}
 We prove lemma 2 by induction on $n$.\\
 \textbf{Base.}
 For $n=5$: level’s 1, 2 nodes of \emph{PQT}$_{5}$ are embedded using the rules specified in table \ref{t3} (figure \ref{fig:n5}).\\
 \begin{table}[H]
\centering
\resizebox{\columnwidth}{!}{%
\begin{tabular}{|c|c|c|c|}
\hline
\emph{Root}   & \emph{Prem(root)} & $0$\emph{suff}$_{1}$ & $0$\emph{pref$_{1}$CQ}$_4$ \\ \hline
$\textcolor{blue}{0}$      & $\textcolor{ForestGreen}{0}\textcolor{red}{00}0000$       & $0\textcolor{blue}{1}$     & $\textcolor{ForestGreen}{0}\textcolor{red}{01}0000$  \\ \hline
$0$\emph{suff}$_{2}$& $1$\emph{pref$_{1}$CQ}$_4$      & $0$\emph{suff}$_{3}$ &$1$\emph{pref$_{0}$CQ}$_4$ \\ \hline
$0\textcolor{blue}{2}$     & $\textcolor{ForestGreen}{1}\textcolor{red}{01}0000$       & $0\textcolor{blue}{3}$     & $\textcolor{ForestGreen}{1}\textcolor{red}{00}0000$  \\ \hline
\end{tabular}
}
\caption{Level’s 1, 2 nodes embedding of \emph{PQT}$_{5}$ into \emph{CQ}$_{7}$.}
\label{t3}
\end{table}

\textbf{Induction hypothesis}

Suppose that for $k \leq n-1$, \emph{PQT}$_{k}$ is one-by-one vertex embedding into \emph{CQ}$_{l}$ with $l < m$ is true.\\
Let us now prove that it is true for $k $=$ n$.

The root $0$ is embedded into $00$\emph{CQ}$_{l-2}$ by using the basic function \emph{Prem(root)}. Nodes of $01$\emph{PQT}$_{k-1}$ are embedded into $0$\emph{CQ}$_{l-1}$ such that: $0$\emph{suff}$_{1}$ is embedded into $0$\emph{pref}$_{1}0000$ of $00$\emph{CQ}$_{l-2}$ using \emph{f}$_{1}$ of situation 2, case 1 (figure \ref{fig:ns2c1}, induction hypothesis). Other nodes of $011$\emph{PQT}$_{k-2}$, $012$\emph{PQT}$_{k-2}$, $013$\emph{PQT}$_{k-2}$ are  respectively embedded into the root embedded component, $01$\emph{CQ}$_{l-2}$ of $0$\emph{CQ}$_{l-1}$, and $01$\emph{CQ}$_{l-2}$ of $0$\emph{CQ}$_{l-1}$ using \emph{f} (figure \ref{fig:ns1}, induction hypothesis).

 Nodes of $02$\emph{PQT}$_{k-1}$ are embedded into $1$\emph{CQ}$_{l-1}$ such that:  $0$\emph{suff}$_{2}$ is embedded into $1$\emph{pref}$_{1}0000$ of $10$\emph{CQ}$_{l-2}$ by definition 02 using \emph{f}$_{1}$ of situation 2, case 1 (figure \ref{fig:ns2c1}). Other nodes of $021$\emph{PQT}$_{k-2}$, $022$\emph{PQT}$_{k-2}$, $023$\emph{PQT}$_{k-2}$ are respectively embedded into the same parent component ($10$\emph{CQ}$_{l-2}$), $11$\emph{CQ}$_{l-2}$ of $1$\emph{CQ}$_{l-1}$, and $11$\emph{CQ}$_{l-2}$ of $1$\emph{CQ}$_{l-1}$ using \emph{f} (figure \ref{fig:ns1}, induction hypothesis).
 
 Nodes of $03$\emph{PQT}$_{k-1}$ are embedded into \emph{CQ}$_{l}$ such that: $0$\emph{suff}$_{3}$ is embedded into $1$\emph{pref}$_{0}0000$ of $10$\emph{CQ}$_{l-2}$ by definition 02 using \emph{f}$_{1}$ of situation 2, case 1 (figure \ref{fig:ns2c1}); and $0$\emph{$3$suff}$_{1}$ is embedded into
this same component $10$\emph{CQ}$_{l-2}$ using \emph{f}$_{1}$ of situation 2, case 2 (figure \ref{fig:ns2c2}, induction hypothesis). Node $0$\emph{$3$suff}$_{2}$ is embedded into $00$\emph{CQ}$_{l-2}$ by definition 02 using \emph{f}$_{1}$ of situation 2, case 2 (figure \ref{fig:ns2c2}). Nodes of $031$\emph{PQT}$_{k-2}$, $032$\emph{PQT}$_{k-2}$ are embedded into the same parent component, respectively $10$\emph{CQ}$_{l-2}$, $00$\emph{CQ}$_{l-2}$ using \emph{f} (figure \ref{fig:ns1}, induction hypothesis).

Nodes of $033$\emph{PQT}$_{k-2}$ are embedded into \emph{CQ}$_{l}$ such that:  $0$\emph{$3$suff}$_{3}$ is embedded into the same parent component ($10$\emph{CQ}$_{l-2}$) using \emph{f}$_{1}$ of situation 2, case 2 (figure \ref{fig:ns2c2}, induction hypothesis). 
For nodes of $0331$\emph{PQT}$_{k-3}$ are embedded into $10$\emph{CQ}$_{l-2}$ and $00$\emph{CQ}$_{l-2}$ such that: 
\begin{itemize}
    \item $0$\emph{$33$suff}$_{1}$ is embedded into the same parent component ($10$\emph{CQ}$_{l-2}$) using \emph{f} (figure \ref{fig:ns1}, induction hypothesis);
    \item $0$\emph{$331$suff}$_{1}$ is embedded into the same parent component ($10$\emph{CQ}$_{l-2}$) using \emph{f}$_{1}$ of situation 2, case 3 (figure \ref{fig:ns2c3}, induction hypothesis);
    \item nodes $0$\emph{$331$suff}$_{2}$ or $0$\emph{$331$suff}$_{3}$ are embedded into $00$\emph{CQ}$_{l-2}$ by definition 02 using \emph{f}$_{1}$ of situation 2, case 3 (figure \ref{fig:ns2c3}).
    
\end{itemize}

For nodes of $0332$\emph{PQT}$_{k-3}$ are embedded into $11$\emph{CQ}$_{l-2}$ and $01$\emph{CQ}$_{l-2}$ such that: 
\begin{itemize}
    \item $0$\emph{$33$suff}$_{2}$ is embedded into $11$\emph{CQ}$_{l-2}$ of $1$\emph{CQ}$_{l-1}$ using \emph{f} (figure \ref{fig:ns1}, induction hypothesis);
    \item nodes $0$\emph{$332$suff}$_{1}$ or $0$\emph{$332$suff}$_{2}$ are embedded into the same parent component ($11$\emph{CQ}$_{l-2}$) using \emph{f}$_{1}$ of situation 2, case 3 (figure \ref{fig:ns2c3}, induction hypothesis);
    \item node $0$\emph{$332$suff}$_{3}$ is embedded into $01$\emph{CQ}$_{l-2}$ of $0$\emph{CQ}$_{l-1}$ by definition 02 using \emph{f}$_{1}$ situation 2, case 3 (figure \ref{fig:ns2c3}).
\end{itemize}

 For nodes of $0333$\emph{PQT}$_{k-3}$ are embedded into $11$\emph{CQ}$_{l-2}$ and $01$\emph{CQ}$_{l-2}$ such that: 
 \begin{itemize}
    \item $0$\emph{$33$suff}$_{3}$ is embedded into $11$\emph{CQ}$_{l-2}$ of $1$\emph{CQ}$_{l-1}$ using \emph{f} (figure \ref{fig:ns1}, induction hypothesis);
    \item nodes  $0$\emph{$333$suff}$_{1}$ or $0$\emph{$333$suff}$_{2}$ are embedded into the same parent component ($11$\emph{CQ}$_{l-2}$) using \emph{f}$_{1}$ of situation 2, case 3 (figure \ref{fig:ns2c3}, induction hypothesis);
    \item node $0$\emph{$333$suff}$_{3}$ is embedded $01$\emph{CQ}$_{l-2}$ of $0$\emph{CQ}$_{l-1}$ by definition 02 using \emph{f}$_{1}$ of situation 2, case 3 (figure \ref{fig:ns2c3}).
\end{itemize}
 
\textbf{Theorem 1.} For $n>5$, a particular sub-quadtree \emph{PQT$_{n}$} is one-by-one vertex embedding into $m$-dimensional crossed cubes \emph{CQ}$_{m}$.\\
\textbf{Proof.}
 We prove theorem 1 by induction on $n$.\\
 \textbf{Base.}
 For $n=6, 8$: level’s 1, 2 nodes of \emph{PQT}$_{6}$, \emph{PQT}$_{8}$ respectively are embedded using the rules specified in table \ref{t4} and table \ref{t5}.\\
 \begin{table}[H]
\centering
\resizebox{\columnwidth}{!}{%
\begin{tabular}{ |c|c|c|c|}
\hline
\emph{Root}   & \emph{Prem(root)} & $0$\emph{suff}$_{1}$ & \emph{pref$_{1}$CQ}$_4$ \\ \hline
$\textcolor{blue}{0}$      & $\textcolor{red}{00000}0000$       & $0\textcolor{blue}{1}$     & $\textcolor{red}{01000}0000$  \\ \hline
$0$\emph{suff}$_{2}$& \emph{pref$_{2}$CQ}$_4$      & $0$\emph{suff}$_{3}$ & \emph{pref$_{3}$CQ}$_4$ \\ \hline
$0\textcolor{blue}{2}$     & $\textcolor{red}{10000}0000$       & $0\textcolor{blue}{3}$     & $\textcolor{red}{11000}0000$  \\ \hline
\end{tabular}
}
\caption{Level’s 1, 2 nodes embedding of \emph{PQT}$_{6}$ into \emph{CQ}$_{9}$.}
\label{t4}
\end{table}

 \begin{table}[H]
\centering
\resizebox{\columnwidth}{!}{%
\begin{tabular}{|c|c|c|c|}
\hline
\emph{Root}   & \emph{Prem(root)} & $0$\emph{suff}$_{1}$ & $0$\emph{pref$_{1}$CQ}$_4$ \\ \hline
$\textcolor{blue}{0}$      & $\textcolor{ForestGreen}{0}\textcolor{red}{0000000}0000$       & $0\textcolor{blue}{1}$     & $\textcolor{ForestGreen}{0}\textcolor{red}{0100000}0000$  \\ \hline
$0$\emph{suff}$_{2}$& $1$\emph{pref$_{1}$CQ}$_4$      & $0$\emph{suff}$_{3}$ &$1$\emph{pref$_{0}$CQ}$_4$ \\ \hline
$0\textcolor{blue}{2}$     & $\textcolor{ForestGreen}{1}\textcolor{red}{0100000}0000$       & $0\textcolor{blue}{3}$     & $\textcolor{ForestGreen}{1}\textcolor{red}{0000000}0000$  \\ \hline
\end{tabular}
}
\caption{Level’s 1, 2 nodes embedding of \emph{PQT}$_{8}$ into \emph{CQ}$_{12}$.}
\label{t5}
\end{table}

 \textbf{Induction hypothesis}
 
 Suppose that for $k \leq n-1$, \emph{PQT}$_{k}$ is one-by-one vertex embedding into \emph{CQ}$_{l}$ with $l < m$ is true.\\
 Let us now prove that it is true for $k = n$.\\
  There are two cases:\\
  \textbf{Case a:} $C=\phi$
  
  One-by-one vertex embedding of $01$\emph{PQT}$_{k-1}$, $02$\emph{PQT}$_{k-1}$, $03$\emph{PQT}$_{k-1}$, and 
  $0$\emph{PQT}$_{k}$ respectively into $01$\emph{CQ}$_{l-2}$, $10$\emph{CQ}$_{l-2}$, $11$\emph{CQ}$_{l-2}$, and $00$\emph{CQ}$_{l-2}$; in this case, we use the same actions as lemma 1.\\
  \textbf{Case b:} $C\neq \phi$
  
  One-by-one vertex embedding of $01$\emph{PQT}$_{k-1}$, $02$\emph{PQT}$_{k-1}$, $03$\emph{PQT}$_{k-1}$, and 
  $0$\emph{PQT}$_{k}$ 
  respectively into $0$\emph{CQ}$_{l-1}$, $1$\emph{CQ}$_{l-1}$, $0$\emph{CQ}$_{l-1}$ and $1$\emph{CQ}$_{l-1}$, and the root $0$ into $00$\emph{CQ}$_{l-2}$; in this case, we use the same actions as lemma 2 except the sub-\emph{PQT}: $0111$\emph{PQT}$_{k-3}$, $0311$\emph{PQT}$_{k-3}$, and $0321$\emph{PQT}$_{k-3}$ are embedded as situation 2, case 1, b as shown in figure \ref{fig:ns2c1}.
  
  \subsection{Dilation two one-by-one edges embedding}
  Dilation two one-by-one edges embedding of \emph{PQT}$_{n}$ onto \emph{CQ}$_{m}$ is done in the following way:\\
  For $n = 3$: the basic function \emph{R} of this dilation two one-by-one edges embedding is produced as follows:
  
\begin{itemize}
\item	\emph{R(A$_{p-1}$-A$_{p-1}$suff$_{1}$) := Pref$_{0}$$00$-Pref$_{1}$$00$} 
\item  \emph{R(A$_{p-1}$-A$_{p-1}$suff$_{2}$) := Pref$_{0}$$00$-Pref$_{2}$$00$ }  
 \item  \emph{R(A$_{p-1}$-A$_{p-1}$suff$_{3}$) := Pref$_{0}$$00$-Pref$_{2}$$00$-Pref$_{3}$$00$}
\end{itemize}

\begin{figure}[H]
\centering
\includegraphics[width=7cm,height=7cm,keepaspectratio]{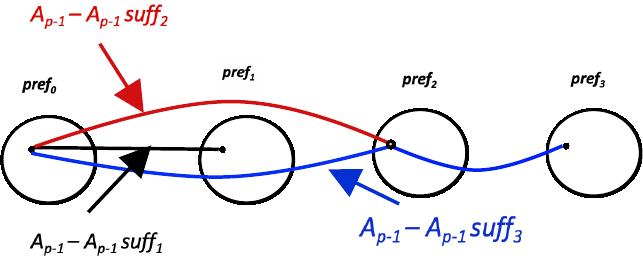}
\caption{Edges embedding situation 1, case 1.} \label{es1c1}
\end{figure}

\begin{figure}[H]
\centering
\includegraphics[width=7cm,height=7cm,keepaspectratio]{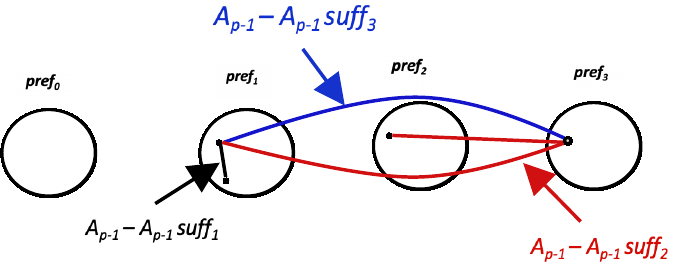}
\caption{Edges embedding situation 1, case 2.} \label{es1c2}
\end{figure}

For $n > 3$: dilation two one-by-one edges embedding is done by two situations. In the first, there are two cases; the first case is when $C = \phi$, we use the basic function \emph{R}; an example is shown in figure \ref{es1c1}.
The second case is when $C \neq \phi$, and we only use one copy $b_{0}$\emph{CQ}$_{m-1}$ or $\bar{b}_{0}$\emph{CQ}$_{m-1}$; the second case is shown in figures \ref{es1c1} and \ref{es1c2}. \emph{R} is produced like situation 1, case 1, or as follows:

\begin{itemize}
\item 	\emph{R(a$_{p-1}$-A$_{p-1}$suff$_{1}$) := Pref$_{1}$$00X$-Pref$_{1}$$00Y$ }
\item  \emph{R(A$_{p-1}$-A$_{p-1}$suff$_{2}$) := Pref$_{1}$$00X$-Pref$_{3}$$00Y$\\-Pref$_{2}$$00Z$}
\item  \emph{R(A$_{p-1}$-A$_{p-1}$suff$_{3}$) := Pref$_{1}$$00X$-Pref$_{3}$$00Y$ }
\end{itemize}

\begin{figure}[H]
\centering
\includegraphics[width=9cm,height=7cm,keepaspectratio]{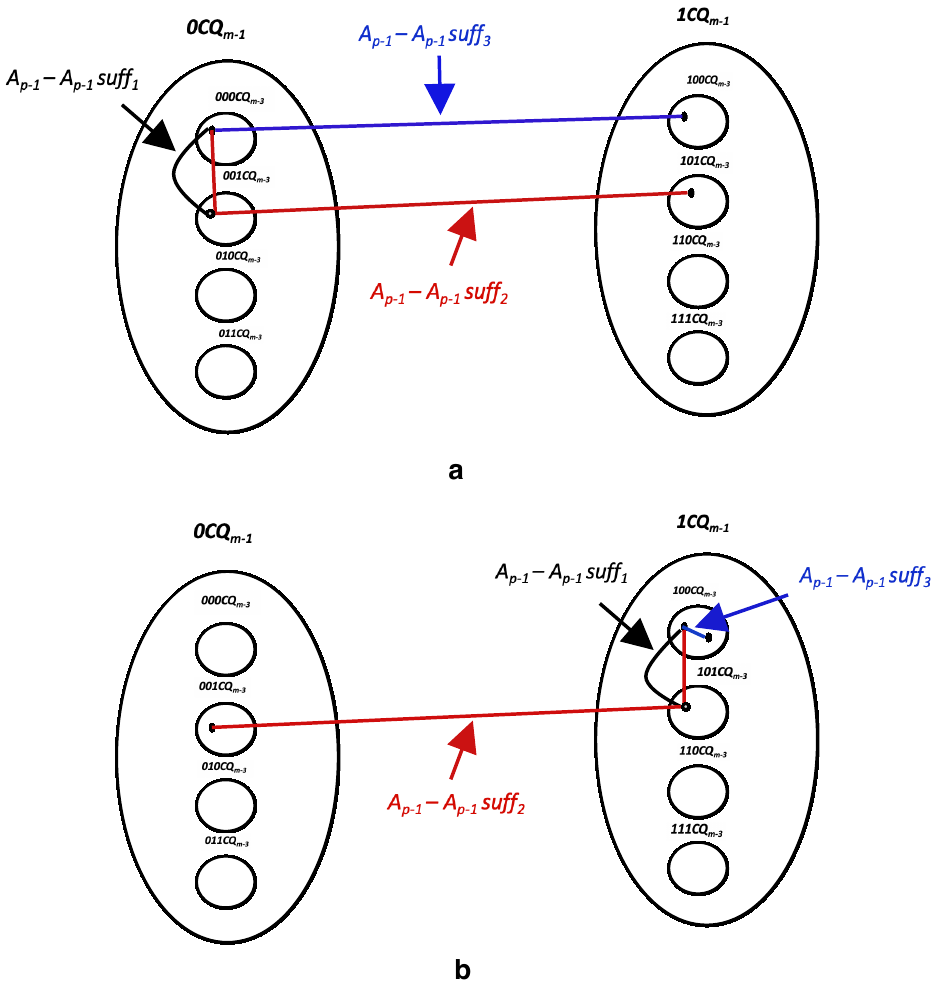}
\caption{Edges embedding situation 2, case 1.} \label{es2c1}
\end{figure}

The second situation is when $C\neq \phi$; in this situation, we use the two copies $b_{0}$\emph{CQ}$_{m-1}$ and $\bar{b}_{0}$\emph{CQ}$_{m-1}$; a function \emph{R}$_{1}$ of this dilation two one-by-one edges embedding. There are three cases; the first is shown in figure \ref{es2c1}. The following rules of case 1 produce \emph{R}$_{1}$:
\begin{itemize}
    \item \emph{R$_{1}$(A$_{p-1}$-A$_{p-1}$suff$_{1}$) := $0$Pref$_{0}$$00X$-$0$Pref$_{1}$$00Y$}
    \item \emph{R$_{1}$(A$_{p-1}$-A$_{p-1}$suff$_{2}$)
    := $0$Pref$_{0}$$00X$-$0$Pref$_{1}$$00Y$\\-$1$Pref$_{1}$$00Z$}
    \item \emph{R$_{1}$(A$_{p-1}$-A$_{p-1}$suff$_{3}$) := $0$Pref$_{0}$$00X$-$1$Pref$_{0}$$00Y$}
\end{itemize}
 \hspace{4cm} OR 
\begin{itemize}
 \item \emph{R$_{1}$(A$_{p-1}$-A$_{p-1}$suff$_{1}$) := $\bar{0}$Pref$_{0}$$00X$-$\bar{0}$Pref$_{1}$$00Y$} 
\item \emph{R$_{1}$(A$_{p-1}$-A$_{p-1}$suff$_{2}$) := $\bar{0}$Pref$_{0}$$00X$-$\bar{0}$Pref$_{1}$$00Y$\\-$\bar{1}$Pref$_{1}$$00Z$} 
\item \emph{R$_{1}$(A$_{p-1}$-A$_{p-1}$suff$_{3}$) := $\bar{0}$Pref$_{0}$$00X$-$1$Pref$_{0}$$00Y$}
\end{itemize}

\begin{figure}[H]
\centering
\includegraphics[width=9cm,height=7cm,keepaspectratio]{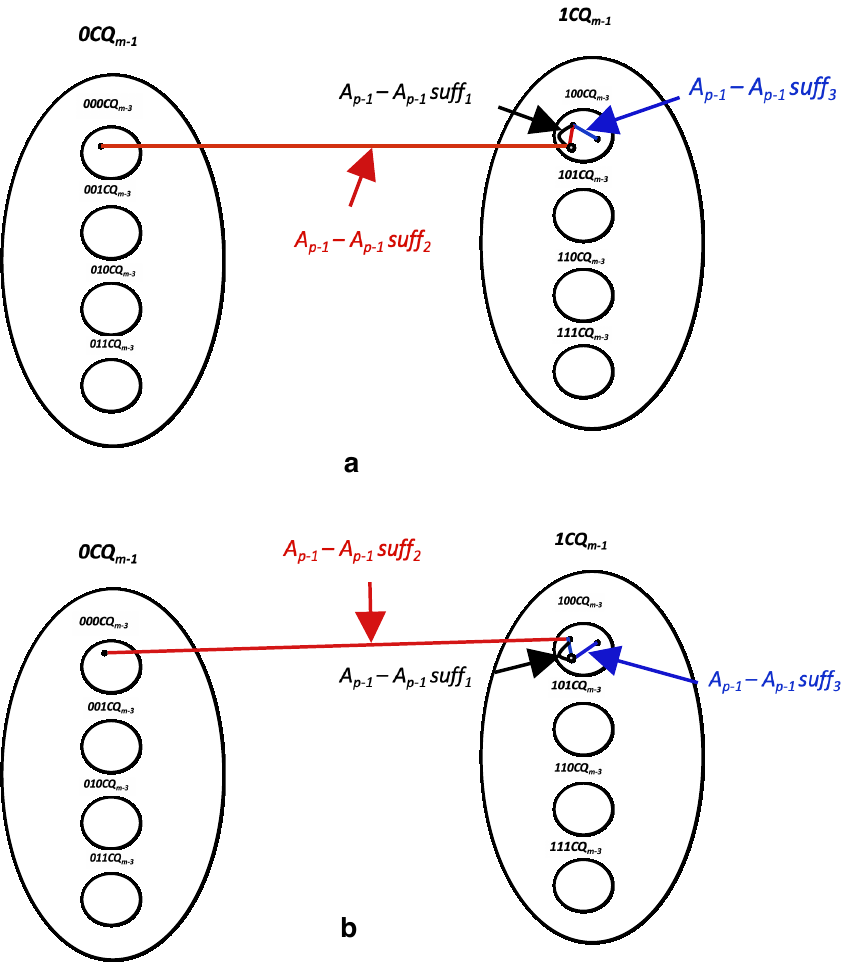}
\caption{Edges embedding situation 2, case 2.} \label{es2c2}
\end{figure}

\begin{figure}[H]
\centering
\includegraphics[width=8cm,height=8cm,keepaspectratio]{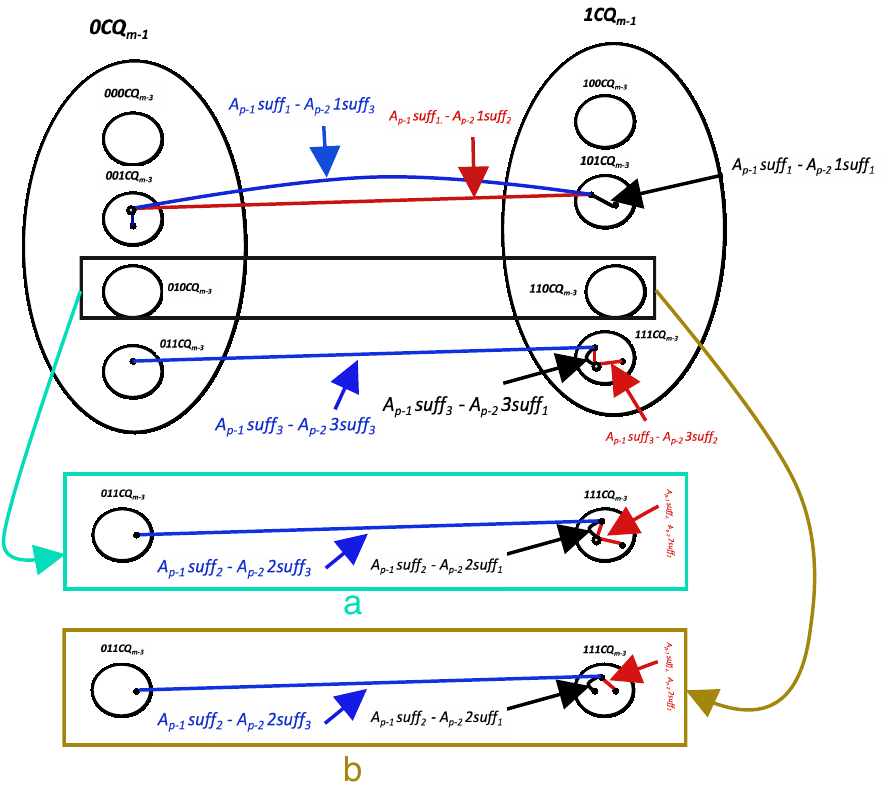}
\caption{Edges embedding situation 2, case 3.} \label{es2c3}
\end{figure}

 The following rules of case 2 shown in figure \ref{es2c2} produce
 \emph{R}$_{1}$:
\begin{itemize}
 \item \emph{R$_{1}$(A$_{p-1}$-A$_{p-1}$suff$_{1}$) := $\bar{0}$Pref$_{0}$$00X$-$\bar{0}$Pref$_{0}$$00Y$} 
 \item  \emph{R$_{1}$(A$_{p-1}$-A$_{p-1}$suff$_{2}$) := $\bar{0}$Pref$_{0}$$00X$-$\bar{0}$Pref$_{0}$$00Y$\\-$\bar{1}$Pref$_{0}$$00Z$} 
 \item  \emph{R$_{1}$(A$_{p-1}$-A$_{p-1}$suff$_{3}$) := $\bar{0}$Pref$_{0}$$00X$-$1$Pref$_{0}$$00Y$}
 \end{itemize}
  \hspace{4cm} OR  
 \begin{itemize}
 \item  \emph{R$_{1}$(A$_{p-1}$-A$_{p-1}$suff$_{1}$) := $\bar{0}$Pref$_{0}$$00X$-$\bar{0}$Pref$_{0}$$00Y$} 
 \item \emph{R$_{1}$(A$_{p-1}$-A$_{p-1}$suff$_{2}$) := $\bar{0}$Pref$_{0}$$00X$-$\bar{1}$Pref$_{0}$$00Y$}
 \item \emph{R$_{1}$(A$_{p-1}$-A$_{p-1}$suff$_{3}$) := $\bar{0}$Pref$_{0}$$00X$-$1$Pref$_{0}$$00Y$\\-$1$Pref$_{0}$$00Z$ }
\end{itemize}
The following rules of the last case shown in figure \ref{es2c3} produce \emph{R}$_{1}$  ($t$ $=$ 2, 3):
\begin{itemize}
\item \emph{R$_{1}$(A$_{p-1}$-A$_{p-1}$suff$_{1}$) := $\bar{0}$Pref$_{1}$$00X$-$\bar{0}$Pref$_{1}$$00Y$}
\item \emph{R$_{1}$(A$_{p-1}$-A$_{p-1}$suff$_{2}$) := $\bar{0}$Pref$_{0+1}$$00X$-$\bar{1}$Pref$_{1}$$00Y$}
\item \emph{R$_{1}$(A$_{p-1}$-A$_{p-1}$suff$_{3}$) := $\bar{0}$Pref$_{1}$$00X$-$0$Pref$_{1}$$00Y$\\-$\bar{1}$Pref$_{1}$$00Z$}

\end{itemize}
 \hspace{4cm} OR 
 \begin{itemize}
 \item \emph{R$_{1}$(A$_{p-1}$-A$_{p-1}$suff$_{1}$) := $\bar{0}$Pref$_{t}$$00X$-\\$\bar{0}$Pref$_{t}$$00Y$}
 \item \emph{R$_{1}$(A$_{p-1}$-A$_{p-1}$suff$_{2}$) := $\bar{0}$Pref$_{t}$$00X$-$\bar{0}$Pref$_{t}$$00Y$-\\$1$Pref$_{1}$$00Z$}
 \item \emph{R$_{1}$(A$_{p-1}$-A$_{p-1}$suff$_{3}$) := $\bar{0}$Pref$_{0}$$00X$-$\bar{1}$Pref$_{t}$$00Y$ }
 \end{itemize}
\hspace{4cm} OR 
 \begin{itemize}
 \item \emph{R$_{1}$(A$_{p-1}$-A$_{p-1}$suff$_{1}$) := $\bar{0}$Pref$_{2}$$00X$-$\bar{0}$Pref$_{2)}$$00Y$}
 \item \emph{R$_{1}$(A$_{p-1}$-A$_{p-1}$suff$_{2}$) := $\bar{0}$Pref$_{2}$$00X$-$1$Pref$_{2}$$00Y$}
 \item \emph{R$_{1}$(A$_{p-1}$-A$_{p-1}$suff$_{3}$) := $\bar{0}$Pref$_{2}$$00X$-$\bar{1}$Pref$_{2}$$00Y$ }
 \end{itemize}

\textbf{Lemma 3.}
For any $n<5$, a particular sub-quadtree \emph{PQT$_{n}$} of dimension \emph{n} is dilation two one-by-one edges embedding onto $m$-dimensional crossed cubes \emph{CQ}$_{m}$.\\
 \textbf{Proof.}
 We prove lemma 3 by induction on $n$.\\
\textbf{Base.}
For $n=\overline{2,4}$: edges between vertices of level 1 and level 2 of \emph{PQT}$_{n}$ are embedded using the rules specified in table \ref{t6}.\\
\begin{table}[H]
\centering
\resizebox{\columnwidth}{!}{%
\begin{tabular}{|c|c|c|c|}
\hline
&\emph{\textbf{PQT}} \textbf{edge} & \textbf{crossed cubes path} & \textbf{Dilation}\\ \hline
\multirow{3}{*}{$n=2$} &\emph{$\textcolor{blue}{0}$}-\emph{$0\textcolor{blue}{1}$}  & \emph{$00$}-\emph{$01$} & 1  \\ \cline{2-4}
&\emph{$\textcolor{blue}{0}$}-\emph{$0\textcolor{blue}{2}$} &\emph{$00$}-\emph{$10$} & 1 \\ \cline{2-4}
 &\emph{$\textcolor{blue}{0}$}-\emph{$0\textcolor{blue}{3}$}   & \emph{$00$}-\emph{$10$}-\emph{$11$} & 2 \\ \hline \hline
 \multirow{3}{*}{$n=3$}& \emph{$\textcolor{blue}{0}$}-\emph{$0\textcolor{blue}{1}$}  & \emph{$\textcolor{red}{00}00$}-\emph{$\textcolor{red}{01}00$} & 1 \\ \cline{2-4}
&\emph{$\textcolor{blue}{0}$}-\emph{$0\textcolor{blue}{2}$} &\emph{$\textcolor{red}{00}00$}-\emph{$\textcolor{red}{10}00$} & 1 \\ \cline{2-4}
 &\emph{$\textcolor{blue}{0}$}-\emph{$0\textcolor{blue}{3}$}   & \emph{$\textcolor{red}{00}00$}-\emph{$\textcolor{red}{10}00$}-\emph{$\textcolor{red}{11}00$} & 2 \\ \hline  \hline
\multirow{3}{*}{$n=4$}& \emph{$\textcolor{blue}{0}$}-\emph{$0\textcolor{blue}{1}$}  & \emph{$\textcolor{red}{00}0000$}-\emph{$\textcolor{red}{01}0000$} & 1 \\ \cline{2-4}
&\emph{$\textcolor{blue}{0}$}-\emph{$0\textcolor{blue}{2}$} &\emph{$\textcolor{red}{00}0000$}-\emph{$\textcolor{red}{10}0000$} & 1 \\ \cline{2-4}
 &\emph{$\textcolor{blue}{0}$}-\emph{$0\textcolor{blue}{3}$}   & \emph{$\textcolor{red}{00}0000$}-\emph{$\textcolor{red}{10}0000$}-\emph{$\textcolor{red}{11}0000$} & 2  \\ \hline 
\end{tabular}
}
\caption{Edges embedding between vertex of level 1 and level 2 of \emph{PQT}$_{2}$, \emph{PQT}$_{3}$, \emph{PQT}$_{4}$.}
\label{t6}
\end{table}


\textbf{Induction hypothesis}

Suppose that for $k \leq n-1$, \emph{PQT}$_{k}$ is dilation two one-by-one edges embedding onto \emph{CQ}$_{l}$ with $l < m$ is true.\\
Is it true for $k = n$ ?.

Edges between the root $0$ and $01$\emph{PQT}$_{k-1}$, $02$\emph{PQT}$_{k-1}$, and $03$\emph{PQT}$_{k-1}$ are embedded respectively onto paths between $00$\emph{CQ}$_{l-2}$ and $01$\emph{CQ}$_{l-2}$, $00$\emph{CQ}$_{l-2}$ and $10$\emph{CQ}$_{l-2}$, and $00$\emph{CQ}$_{l-2}$ and $10$\emph{CQ}$_{l-2}$ and $11$\emph{CQ}$_{l-2}$. Edges in this lemma are embedded using \emph{R} (induction hypothesis, situation 1, case 1, as shown in figure \ref{es1c1}).\\

\textbf{Lemma 4.}
 For any $n \geq 5$, all edges in the same sub-\emph{PQT$_{5}$} of any \emph{PQT$_{n}$} is dilation two one-by-one edges embedding onto paths in the same sub-\emph{CQ}$_{7}$ of any supernode of \emph{CQ}$_{m}$.\\
 \textbf{Proof.}
 We prove lemma 4 by induction on $n$.\\
\textbf{Base.}
For $n=5$: edges of \emph{PQT}$_{5}$ are embedded using the rules specified in table \ref{t7}.
\begin{table}[H]
\centering
\resizebox{\columnwidth}{!}{%
\begin{tabular}{|c|c|c|c|}
\hline
&\emph{\textbf{PQT}} \textbf{edge} & \textbf{crossed cubes path} & \textbf{Dilation} \\ \hline
 \multirow{3}{*}{A}&\emph{$\textcolor{blue}{0}$}-\emph{$0\textcolor{blue}{1}$}  & \emph{$\textcolor{ForestGreen}{0}\textcolor{red}{00}0000$}-\emph{$\textcolor{ForestGreen}{0}\textcolor{red}{01}0000$} & 1 \\ \cline{2-4}
&\emph{$\textcolor{blue}{0}$}-\emph{$0\textcolor{blue}{2}$} &\emph{$\textcolor{ForestGreen}{0}\textcolor{red}{00}0000$}-\emph{$\textcolor{ForestGreen}{0}\textcolor{red}{01}0000$}-\emph{$\textcolor{ForestGreen}{1}\textcolor{red}{01}0000$} & 2 \\ \cline{2-4}
 &\emph{$\textcolor{blue}{0}$}-\emph{$0\textcolor{blue}{3}$}   & \emph{$\textcolor{ForestGreen}{0}\textcolor{red}{00}0000$}-\emph{$\textcolor{ForestGreen}{1}\textcolor{red}{00}0000$} & 1  \\ \hline \hline
 \multirow{3}{*}{B}& \emph{$0\textcolor{blue}{1}$}-\emph{$01\textcolor{blue}{1}$}  & \emph{$\textcolor{ForestGreen}{0}\textcolor{red}{01}0000$}-\emph{$\textcolor{ForestGreen}{0}\textcolor{red}{01}0001$} & 1  \\ \cline{2-4}
&\emph{$0\textcolor{blue}{1}$}-\emph{$01\textcolor{blue}{2}$} &\emph{$\textcolor{ForestGreen}{0}\textcolor{red}{01}0000$}-\emph{$\textcolor{ForestGreen}{0}\textcolor{red}{11}0000$}-\emph{$\textcolor{ForestGreen}{0}\textcolor{red}{10}0000$} & 2 \\ \cline{2-4}
 &\emph{$0\textcolor{blue}{1}$}-\emph{$01\textcolor{blue}{3}$}   & \emph{$\textcolor{ForestGreen}{0}\textcolor{red}{01}0010$}-\emph{$\textcolor{ForestGreen}{0}\textcolor{red}{11}0000$} & 1 \\ \hline \hline
  \multirow{3}{*}{C}&\emph{$0\textcolor{blue}{3}$}-\emph{$03\textcolor{blue}{1}$}  & \emph{$\textcolor{ForestGreen}{1}\textcolor{red}{00}0000$}-\emph{$\textcolor{ForestGreen}{1}\textcolor{red}{00}0001$} & 1  \\ \cline{2-4}
&\emph{$0\textcolor{blue}{3}$}-\emph{$03\textcolor{blue}{2}$} &\emph{$\textcolor{ForestGreen}{1}\textcolor{red}{00}0000$}-\emph{$\textcolor{ForestGreen}{1}\textcolor{red}{00}0001$}-\emph{$\textcolor{ForestGreen}{0}\textcolor{red}{00}0011$} & 2 \\ \cline{2-4}
 &\emph{$0\textcolor{blue}{3}$}-\emph{$03\textcolor{blue}{3}$}   & \emph{$\textcolor{ForestGreen}{1}\textcolor{red}{00}0000$}-\emph{$\textcolor{ForestGreen}{1}\textcolor{red}{00}0010$} & 1  \\ \hline \hline
\multirow{3}{*}{D} &\emph{$033\textcolor{blue}{1}$}-\emph{$0331\textcolor{blue}{1}$}  & \emph{$\textcolor{ForestGreen}{1}\textcolor{red}{01}0010$}-\emph{$\textcolor{ForestGreen}{1}\textcolor{red}{01}0011$} & 1 \\ \cline{2-4}
&\emph{$033\textcolor{blue}{1}$}-\emph{$0331\textcolor{blue}{2}$} &\emph{$\textcolor{ForestGreen}{1}\textcolor{red}{01}0010$}-\emph{$\textcolor{ForestGreen}{0}\textcolor{red}{01}0010$}& 1 \\ \cline{2-4}
 &\emph{$033\textcolor{blue}{1}$}-\emph{$0331\textcolor{blue}{3}$}   & \emph{$\textcolor{ForestGreen}{1}\textcolor{red}{01}0010$}-\emph{$\textcolor{ForestGreen}{0}\textcolor{red}{01}0010$}-\emph{$\textcolor{ForestGreen}{0}\textcolor{red}{01}0011$} & 2 \\ \hline
\end{tabular}
}
\caption{Edges embedding of \emph{PQT}$_{5}$ into \emph{CQ}$_{7}$, A: example of situation 2, case 1; B: example of situation 1, case 2; C: example of situation 2, case 2; D: example of situation 2, case 3.}
\label{t7}
\end{table}
\textbf{Induction hypothesis}

Suppose that for $k \leq n-1$, any sub-\emph{PQT$_{5}$} of \emph{PQT$_{k}$} is dilation two one-by-one edges embedding onto any sub-\emph{CQ}$_{7}$ of \emph{CQ}$_{l}$ with $l < m$ is true.\\
Is it true for $k = n$ ?.

For any sub-\emph{PQT}$_{k^{'}}$, sub-\emph{CQ}$_{l^{'}}$ with $k^{'} = 5$, $l^{'}= 7$; the edge between $0$\emph{PQT}$_{k^{'}}$, $01$\emph{PQT}$_{k^{'}-1}$ is embedded onto a path in the same component $00$\emph{CQ}$_{l^{'}-2}$ using \emph{R}$_{1}$ of situation 2, case 1 (figure \ref{es2c1}, induction hypothesis). For edges of $01$\emph{PQT}$_{k^{'}-1}$ are embedded onto paths between $00$\emph{CQ}$_{l^{'}-2}$, $01$\emph{CQ}$_{l^{'}-2}$, and the same $00$\emph{CQ}$_{l^{'}-2}$, $01$\emph{CQ}$_{l^{'}-2}$ of $0$\emph{CQ}$_{l^{'}-1}$ using \emph{R} (figures \ref{es1c1}, \ref{es1c2}, induction hypothesis).

 The edge between $0$\emph{PQT}$_{k^{'}}$, $02$\emph{PQT}$_{k^{'}-1}$ is embedded onto a path in the same component $00$\emph{CQ}$_{l^{'}-2}$ (induction hypothesis), and between $00$\emph{CQ}$_{l^{'}-2}$, $10$\emph{CQ}$_{l^{'}-2}$ by definition 02  using \emph{R}$_{1}$ of situation 2, case 1 (figure \ref{es2c1}). For edges of $02$\emph{PQT}$_{k^{'}-1}$ are embedded onto paths between $10$\emph{CQ}$_{l^{'}-2}$, $11$\emph{CQ}$_{l^{'}-2}$, and the same $10$\emph{CQ}$_{l^{'}-2}$, $11$\emph{CQ}$_{l^{'}-2}$ of $1$\emph{CQ}$_{l^{'}-1}$ using \emph{R} (figures \ref{es1c1}, \ref{es1c2}, induction hypothesis).
 
 The edge between $0$\emph{PQT}$_{k^{'}}$, $03$\emph{PQT}$_{k^{'}-1}$ is embedded onto a path between $00$\emph{CQ}$_{l^{'}-2}$, $10$\emph{CQ}$_{l^{'}-2}$ by definition 02  using \emph{R}$_{1}$ of situation 2, case 1 (figure \ref{es2c1}). For edges of $03$\emph{PQT}$_{k^{'}-1}$ are embedded as follows: the edge between $03$\emph{PQT}$_{k^{'}-1}$, $031$\emph{PQT}$_{k^{'}-2}$ is embedded onto a path in the same component $10$\emph{CQ}$_{l^{'}-2}$ using \emph{R}$_{1}$ of situation 2, case 2
 (figure \ref{es2c2}, induction hypothesis); the edge between $03$\emph{PQT}$_{k^{'}-1}$, $032$\emph{PQT}$_{k^{'}-2}$ is embedded onto a path in the same component $10$\emph{CQ}$_{l^{'}-2}$ (induction hypothesis), and between $10$\emph{CQ}$_{l^{'}-2}$, $00$\emph{CQ}$_{l^{'}-2}$ by definition 02  using \emph{R}$_{1}$ of situation 2, case 2 (figure \ref{es2c2}). Edges of $031$\emph{PQT}$_{k^{'}-2}$, $032$\emph{PQT}$_{k^{'}-2}$ are embedded onto paths in the same $00$\emph{CQ}$_{l^{'}-2}$, $10$\emph{CQ}$_{l^{'}-2}$ using \emph{R} of situation 1, case 1(figure \ref{es1c1}, induction hypothesis).
 
 The edge between $03$\emph{PQT}$_{k^{'}-1}$, $033$\emph{PQT}$_{k^{'}-2}$ is embedded onto a path in the same component $10$\emph{CQ}$_{l^{'}-2}$ using \emph{R}$_{1}$ of situation 2, case 2 (figure \ref{es2c2}, induction hypothesis). For edges of $033$\emph{PQT}$_{k^{'}-2}$ are embedded as follows:  the edge between $033$\emph{PQT}$_{k^{'}-2}$, $0331$\emph{PQT}$_{k^{'}-3}$ is embedded onto a path in the same component $10$\emph{CQ}$_{l^{'}-2}$ using \emph{R} of situation 1, case 1 (figure \ref{es1c1}, induction hypothesis). Edges of $0331$\emph{PQT}$_{k^{'}-3}$ are embedded onto paths in the same component $10$\emph{CQ}$_{l^{'}-2}$ (induction hypothesis), between $10$\emph{CQ}$_{l^{'}-2}$, $00$\emph{CQ}$_{l^{'}-2}$ by definition 02 , and in the same component $00$\emph{CQ}$_{l^{'}-2}$ (induction hypothesis) using \emph{R}$_{1}$ of situation 2, case 3 (figure \ref{es2c3}). 
 
 Edges between $033$\emph{PQT}$_{k^{'}-2}$, $0332$\emph{PQT}$_{k^{'}-3}$ or $0333$\emph{PQT}$_{k^{'}-3}$ are embedded onto paths between $10$\emph{CQ}$_{l^{'}-2}$, $11$\emph{CQ}$_{l^{'}-2}$ of $1$\emph{CQ}$_{l^{'}-1}$ using \emph{R} of situation 1, case 1 (figure \ref{es1c1}, induction hypothesis). 
 
 Edges of $0332$\emph{PQT}$_{k^{'}-3}$ or $0333$\emph{PQT}$_{k^{'}-3}$ are embedded onto paths in the same component $11$\emph{CQ}$_{l^{'}-2}$ (induction hypothesis), and between $11$\emph{CQ}$_{l^{'}-2}$, $01$\emph{CQ}$_{l^{'}-2}$ by definition 02  using \emph{R}$_{1}$ of situation 2, case 3 (figure \ref{es2c3}).\\
 
 \textbf{Theorem 2.} For any $n > 5$, a particular sub-quadtree \emph{PQT$_{n}$} is dilation two one-by-one edges embedding onto $m$-dimensional crossed cubes \emph{CQ}$_{m}$.\\
\textbf{Proof.}
 We prove theorem 2 by induction on $n$.\\
 \textbf{Base.}
 For $n=6, 8$: edges of \emph{PQT}$_{6}$, \emph{PQT}$_{8}$ are respectively embedded using the rules specified in table \ref{t8}, table \ref{t9}.
  \begin{table}[H]
\centering
\resizebox{\columnwidth}{!}{%
\begin{tabular}{|c|c|c|}
\hline
\emph{\textbf{PQT}} \textbf{edge} & \textbf{crossed cubes path} & \textbf{Dilation} \\ \hline
 \emph{$\textcolor{blue}{0}$}-\emph{$0\textcolor{blue}{1}$}  & \emph{$\textcolor{red}{00000}0000$}-\emph{$\textcolor{red}{01000}0000$} & 1  \\ \hline
\emph{$\textcolor{blue}{0}$}-\emph{$0\textcolor{blue}{2}$} &\emph{$\textcolor{red}{00000}0000$}-\emph{$\textcolor{red}{10000}0000$} & 1 \\ \hline
 \emph{$\textcolor{blue}{0}$}-\emph{$0\textcolor{blue}{3}$}   & \emph{$\textcolor{red}{00000}0000$}-\emph{$\textcolor{red}{10000}0000$}-\emph{$\textcolor{red}{11000}0000$} & 2  \\ \hline
\end{tabular}
}
\caption{Edges embedding of \emph{PQT}$_{6}$ into \emph{CQ}$_{9}$.}
\label{t8}
\end{table}

 \textbf{Induction hypothesis}
 
 Suppose that for $k \leq n-1$, \emph{PQT}$_{k}$ is dilation two one-by-one edges embedding onto \emph{CQ}$_{l}$ with $l < m$ is true.\\
 Is it true for $k = n$ ?.\\
  There are two cases:\\
  \textbf{Case a:} $C =\phi$
  
   Dilation two one-by-one edges embedding between the root $0$ and respectively $01$\emph{PQT}$_{k-1}$, $02$\emph{PQT}$_{k-1}$, and $03$\emph{PQT}$_{k-1}$
   onto paths respectively between $00$\emph{CQ}$_{l-2}$ and $01$\emph{CQ}$_{l-2}$, $00$\emph{CQ}$_{l-2}$ and $10$\emph{CQ}$_{l-2}$, $00$\emph{CQ}$_{l-2}$ and $11$\emph{CQ}$_{l-2}$. In this case, we use the same actions as lemma 3.\\
  \textbf{Case b:} $C\neq  \phi$
  
  Dilation two one-by-one edges embedding between the root $0$ and respectively $01$\emph{PQT}$_{k-1}$, $02$\emph{PQT}$_{k-1}$, and $03$\emph{PQT}$_{k-1}$
   onto paths in the same supernode $00$\emph{CQ}$_{l-2}$, and between $00$\emph{CQ}$_{l-2}$, $10$\emph{CQ}$_{l-2}$ (situation 2, case 1, a figure \ref{es2c1}). Dilation two one-by-one edges of $01$\emph{PQT}$_{k-1}$, $02$\emph{PQT}$_{k-1}$, and $03$\emph{PQT}$_{k-1}$ 
 respectively onto paths in $0$\emph{CQ}$_{l-1}$, $1$\emph{CQ}$_{l-1}$, and both $0$\emph{CQ}$_{l-1}$, $1$\emph{CQ}$_{l-1}$.
 
 In this case, we use the same actions as lemma 4 except the edges of sub-\emph{PQT}: $0111$\emph{PQT}$_{k-3}$, $0311$\emph{PQT}$_{k-3}$, and $0321$\emph{PQT}$_{k-3}$ are embedded like situation 2, case 1, b (figure \ref{es2c1}). Moreover, edges of sub-\emph{PQT}: $01113$\emph{PQT}$_{k-4}$, $03113$\emph{PQT}$_{k-4}$, and $03213$\emph{PQT}$_{k-4}$ are embedded as situation 2, case 2, b (figure \ref{es2c2}).
 
 \begin{table}[H]
\centering
\resizebox{\columnwidth}{!}{%

\begin{tabular}{|c|c|c|c|}
\hline
& \emph{\textbf{PQT}} \textbf{edge} & \textbf{crossed cubes path} & \textbf{Dilation} \\ \hline
\multirow{3}{*}{A} &\shortstack{\vspace{0.15cm}\emph{$\textcolor{blue}{0}$}-\emph{$0\textcolor{blue}{1}$} } &\shortstack{\\\emph{$\textcolor{ForestGreen}{0}\textcolor{red}{0000000}0000$}\\-\emph{$\textcolor{ForestGreen}{0}\textcolor{red}{0100000}0000$}} & \shortstack{\vspace{0.15cm}1} \\ \cline{2-4}
&\shortstack{\vspace{0.15cm}\emph{$\textcolor{blue}{0}$}-\emph{$0\textcolor{blue}{2}$}} &\shortstack{\\\emph{$\textcolor{ForestGreen}{0}\textcolor{red}{0000000}0000$}-\emph{$\textcolor{ForestGreen}{0}\textcolor{red}{0100000}0000$}\\-\emph{$\textcolor{ForestGreen}{1}\textcolor{red}{0100000}0000$}} &\shortstack{\vspace{0.15cm} 2}\\ \cline{2-4}
& \shortstack{\vspace{0.15cm}\emph{$\textcolor{blue}{0}$}-\emph{$0\textcolor{blue}{3}$}}   & \shortstack{\\\emph{$\textcolor{ForestGreen}{0}\textcolor{red}{0000000}0000$}\\-\emph{$\textcolor{ForestGreen}{1}\textcolor{red}{0000000}0000$}} &\shortstack{\vspace{0.15cm} 1 } \\ \hline \hline
 \multirow{3}{*}{B}&\shortstack{\vspace{0.15cm}\emph{$0\textcolor{blue}{3}$}-\emph{$03\textcolor{blue}{1}$}}  & \shortstack{\\\emph{$\textcolor{ForestGreen}{1}\textcolor{red}{0000000}0000$}\\-\emph{$\textcolor{ForestGreen}{1}\textcolor{red}{0001000}0000$}} & \shortstack{\vspace{0.15cm}1}  \\ \cline{2-4}
&\shortstack{\vspace{0.15cm}\emph{$0\textcolor{blue}{3}$}-\emph{$03\textcolor{blue}{2}$}} &\shortstack{\\\emph{$\textcolor{ForestGreen}{1}\textcolor{red}{0000000}0000$}-\\\emph{$\textcolor{ForestGreen}{1}\textcolor{red}{0001000}0000$}-\emph{$\textcolor{ForestGreen}{0}\textcolor{red}{0001000}0000$}}&\shortstack{\vspace{0.15cm} 2 } \\ \cline{2-4}
 &\shortstack{\vspace{0.15cm}\emph{$0\textcolor{blue}{3}$}-\emph{$03\textcolor{blue}{3}$} }  & \shortstack{\\\emph{$\textcolor{ForestGreen}{1}\textcolor{red}{0000000}0000$}\\-\emph{$\textcolor{ForestGreen}{1}\textcolor{red}{0000000}0001$}} &\shortstack{\vspace{0.15cm} 1 } \\ \hline \hline
 \multirow{3}{*}{C} & \shortstack{\vspace{0.15cm}\emph{$0\textcolor{blue}{1}$}-\emph{$01\textcolor{blue}{1}$} } & \shortstack{\\\emph{$\textcolor{Orange}{0}\textcolor{red}{0100000}0000$}\\-\emph{$\textcolor{Orange}{0}\textcolor{red}{0101000}0000$}} &\shortstack{\vspace{0.15cm} 1 } \\ \cline{2-4}
&\shortstack{\vspace{0.15cm}\emph{$0\textcolor{blue}{1}$}-\emph{$01\textcolor{blue}{2}$}} &\shortstack{\\\emph{$\textcolor{Orange}{0}\textcolor{red}{0100000}0000$}-\\\emph{$\textcolor{Orange}{0}\textcolor{red}{1100000}0000$}-\emph{$\textcolor{Orange}{0}\textcolor{red}{1000000}0000$}} &\shortstack{\vspace{0.15cm} 2} \\ \cline{2-4}
& \shortstack{\vspace{0.15cm}\emph{$0\textcolor{blue}{1}$}-\emph{$01\textcolor{blue}{3}$} }  & \shortstack{\\\emph{$\textcolor{Orange}{0}\textcolor{red}{0100000}0000$}\\-\emph{$\textcolor{Orange}{0}\textcolor{red}{1100000}0000$}} & \shortstack{\vspace{0.15cm}1 } \\ \hline \hline
\multirow{3}{*}{D} & \shortstack{\vspace{0.15cm}\emph{$01\textcolor{blue}{1}$}-\emph{$011\textcolor{blue}{1}$} } & \shortstack{\\\emph{$\textcolor{Orange}{001}\textcolor{red}{01000}0000$}\\-\emph{$\textcolor{Orange}{001}\textcolor{red}{01100}0000$}} & \shortstack{\vspace{0.15cm}1 } \\ \cline{2-4}
&\shortstack{\vspace{0.15cm}\emph{$01\textcolor{blue}{1}$}-\emph{$011\textcolor{blue}{2}$}} &\shortstack{\\\emph{$\textcolor{Orange}{001}\textcolor{red}{01000}0000$}-\\\emph{$\textcolor{Orange}{001}\textcolor{red}{11000}0000$}-\emph{$\textcolor{Orange}{001}\textcolor{red}{10000}0000$}} &\shortstack{\vspace{0.15cm} 2 }\\ \cline{2-4}
 &\shortstack{\vspace{0.15cm}\emph{$01\textcolor{blue}{1}$}-\emph{$011\textcolor{blue}{3}$} }  & \shortstack{\\\emph{$\textcolor{Orange}{001}\textcolor{red}{01000}0000$}\\-\emph{$\textcolor{Orange}{001}\textcolor{red}{11000}0000$}} & \shortstack{\vspace{0.15cm} 1}  \\ \hline \hline
 \multirow{3}{*}{E}&\shortstack{\vspace{0.15cm}\emph{$011\textcolor{blue}{1}$}-\emph{$0111\textcolor{blue}{1}$}}  & \shortstack{\\\emph{$\textcolor{Orange}{00101}\textcolor{ForestGreen}{1}\textcolor{red}{00}0000$}\\-\emph{$\textcolor{Orange}{00101}\textcolor{ForestGreen}{1}\textcolor{red}{01}0000$}} & \shortstack{\vspace{0.15cm}1} \\ \cline{2-4}
&\shortstack{\vspace{0.15cm}\emph{$011\textcolor{blue}{1}$}-\emph{$0111\textcolor{blue}{2}$}} &\shortstack{\\\emph{$\textcolor{Orange}{00101}\textcolor{ForestGreen}{1}\textcolor{red}{00}0000$}-\\\emph{$\textcolor{Orange}{00101}\textcolor{ForestGreen}{1}\textcolor{red}{01}0000$}-\emph{$\textcolor{Orange}{00101}\textcolor{ForestGreen}{0}\textcolor{red}{01}0000$}} & \shortstack{\vspace{0.15cm}2} \\ \cline{2-4}
 &\shortstack{\vspace{0.15cm}\emph{$011\textcolor{blue}{1}$}-\emph{$0111\textcolor{blue}{3}$} }  & \shortstack{\\\emph{$\textcolor{Orange}{00101}\textcolor{ForestGreen}{1}\textcolor{red}{00}0000$}\\-\emph{$\textcolor{Orange}{00101}\textcolor{ForestGreen}{1}\textcolor{red}{00}0001$}} & \shortstack{\vspace{0.15cm}1 } \\ \hline
\end{tabular}
}
\caption{Edges embedding of \emph{PQT}$_{8}$ into \emph{CQ}$_{12}$, A: example of situation 2, case 1; B: example of situation 2, case 2; C, D: example of situation 1, case 2; E: example of situation 2, case 1.}
\label{t9}
\end{table}
\section{Conclusion}
In this paper, we have proposed a new function for embedding \emph{PQT}$_{n}$ into  \emph{CQ}$_{m}$. The main purpose is dilation two one-by-one embedding \emph{PQT}$_{n}$ into  \emph{CQ}$_{m}$. The study of dilation of this function is explained in three steps. The first step is dilation two one-by-one embeddings all edges onto paths in the same \emph{CQ}$_{4}$ of any supernode of \emph{CQ}$_{m}$ as proved by lemma 3. The second step is dilation two one-by-one embeddings all edges onto paths in the same \emph{CQ}$_{7}$ of any supernode of \emph{CQ}$_{m}$ as proved by lemma 4. The third step is the general dilation two one-by-one embeddings all edges of \emph{PQT}$_{n}$ onto paths between two supernodes of \emph{CQ}$_{m}$ as proved by theorem 2.\\
As a perspective, it is more interesting to study the fault-tolerant embedding of \emph{PQT}$_{n}$ into \emph{CQ}$_{m}$.

\section*{CRediT authorship contribution statement}
\textbf{Selmi Aymen Takie eddine:} ConceptuaIization, Writing- original draft, Writing-review \& editing, Investigation. \textbf{Mohamed Faouzi Zerarka:} Conceptualization, Writing–review \& editing. \textbf{Abdelhakim Cheriet:}  Conceptualization.

\section*{Declaration of competing interest}
The authors declare that they have no known competing financial interests or personal relationships that could have appeared to influence the work reported in this paper.

\section*{Acknowledgment}
This work was funded in part by the Algerian Ministry of Higher Education and Scientific Research under contrat PRFU C00L07UN070120210004.

\bibliographystyle{model1-num-names}
\bibliography{ref.bib}

\begin{thebibliography}{33}
\expandafter\ifx\csname natexlab\endcsname\relax\def\natexlab#1{#1}\fi
\providecommand{\bibinfo}[2]{#2}
\ifx\xfnm\relax \def\xfnm[#1]{\unskip,\space#1}\fi
\bibitem[{Zhou et~al.(2020)Zhou, Ben, Wang, Zheng, and Du}]{14}
\bibinfo{author}{J.~Zhou}, \bibinfo{author}{J.~Ben}, \bibinfo{author}{R.~Wang},
  \bibinfo{author}{M.~Zheng}, \bibinfo{author}{L.~Du},
\newblock \bibinfo{title}{Lattice quad-tree indexing algorithm for a hexagonal
  discrete global grid system},
\newblock \bibinfo{journal}{ISPRS International Journal of Geo-Information}
  \bibinfo{volume}{9} (\bibinfo{year}{2020}) \bibinfo{pages}{83}.
\bibitem[{Shukla et~al.(2020)Shukla, Verma, Verma, Kumar et~al.}]{15}
\bibinfo{author}{P.~Shukla}, \bibinfo{author}{A.~Verma},
  \bibinfo{author}{S.~Verma}, \bibinfo{author}{M.~Kumar}, et~al.,
\newblock \bibinfo{title}{Interpreting svm for medical images using quadtree},
\newblock \bibinfo{journal}{Multimedia Tools and Applications}
  \bibinfo{volume}{79} (\bibinfo{year}{2020}) \bibinfo{pages}{29353--29373}.
\bibitem[{Banerjee et~al.(2020)Banerjee, Wang, Chopp, Cossairt, and
  Katsaggelos}]{17}
\bibinfo{author}{S.~Banerjee}, \bibinfo{author}{Z.~W. Wang},
  \bibinfo{author}{H.~H. Chopp}, \bibinfo{author}{O.~Cossairt},
  \bibinfo{author}{A.~Katsaggelos},
\newblock \bibinfo{title}{Quadtree driven lossy event compression},
\newblock \bibinfo{journal}{arXiv preprint arXiv:2005.00974}
  (\bibinfo{year}{2020}).
\bibitem[{Shen and Zhao(2020)}]{29}
\bibinfo{author}{Q.~Shen}, \bibinfo{author}{Y.~Zhao},
\newblock \bibinfo{title}{Perceptual hashing for color image based on color
  opponent component and quadtree structure},
\newblock \bibinfo{journal}{Signal Processing} \bibinfo{volume}{166}
  (\bibinfo{year}{2020}) \bibinfo{pages}{107244}.
\bibitem[{Albani et~al.(2021)Albani, H{\"o}nig, Nardi, Ayanian, and
  Trianni}]{30}
\bibinfo{author}{D.~Albani}, \bibinfo{author}{W.~H{\"o}nig},
  \bibinfo{author}{D.~Nardi}, \bibinfo{author}{N.~Ayanian},
  \bibinfo{author}{V.~Trianni},
\newblock \bibinfo{title}{Hierarchical task assignment and path finding with
  limited communication for robot swarms},
\newblock \bibinfo{journal}{Applied Sciences} \bibinfo{volume}{11}
  (\bibinfo{year}{2021}) \bibinfo{pages}{3115}.
\bibitem[{Efe and Fern{\'a}ndez(1996)}]{23}
\bibinfo{author}{K.~Efe}, \bibinfo{author}{A.~Fern{\'a}ndez},
\newblock \bibinfo{title}{Mesh-connected trees: a bridge between grids and
  meshes of trees},
\newblock \bibinfo{journal}{IEEE Transactions on Parallel and Distributed
  Systems} \bibinfo{volume}{7} (\bibinfo{year}{1996})
  \bibinfo{pages}{1281--1291}.
\bibitem[{Rabie and Kamel(2017)}]{31}
\bibinfo{author}{T.~Rabie}, \bibinfo{author}{I.~Kamel},
\newblock \bibinfo{title}{Toward optimal embedding capacity for transform
  domain steganography: a quad-tree adaptive-region approach},
\newblock \bibinfo{journal}{Multimedia Tools and Applications}
  \bibinfo{volume}{76} (\bibinfo{year}{2017}) \bibinfo{pages}{8627--8650}.
\bibitem[{Jaillet and Lobos(2021)}]{32}
\bibinfo{author}{F.~Jaillet}, \bibinfo{author}{C.~Lobos},
\newblock \bibinfo{title}{Fast quadtree/octree adaptive meshing and re-meshing
  with linear mixed elements},
\newblock \bibinfo{journal}{Engineering with Computers}  (\bibinfo{year}{2021})
  \bibinfo{pages}{1--18}.
\bibitem[{Gupta et~al.(2003)Gupta, Nelson, and Wang}]{33}
\bibinfo{author}{A.~K. Gupta}, \bibinfo{author}{D.~Nelson},
  \bibinfo{author}{H.~Wang},
\newblock \bibinfo{title}{Efficient embeddings of ternary trees into
  hypercubes},
\newblock \bibinfo{journal}{Journal of Parallel and Distributed Computing}
  \bibinfo{volume}{63} (\bibinfo{year}{2003}) \bibinfo{pages}{619--629}.
\bibitem[{Abuelrub(2010)}]{9}
\bibinfo{author}{E.~Abuelrub},
\newblock \bibinfo{title}{Embedding interconnection networks in crossed cubes},
\newblock in: \bibinfo{booktitle}{Electronic Engineering and Computing
  Technology}, \bibinfo{publisher}{Springer}, \bibinfo{year}{2010}, pp.
  \bibinfo{pages}{141--151}.
\bibitem[{Leighton(1991)}]{5}
\bibinfo{author}{F.~T. Leighton}, \bibinfo{title}{Introduction to Parallel
  Algorithms and Architectures: Array, Trees, Hypercubes},
  \bibinfo{publisher}{Morgan Kaufmann Publishers Inc.}, \bibinfo{address}{San
  Francisco, CA, USA}, \bibinfo{year}{1991}.
\bibitem[{Barasch et~al.(1989)Barasch, Lakshmivarahan, and Dhall}]{6}
\bibinfo{author}{L.~Barasch}, \bibinfo{author}{S.~Lakshmivarahan},
  \bibinfo{author}{S.~Dhall},
\newblock \bibinfo{title}{Embedding arbitrary meshes and complete binary trees
  in generalized hypercubes},
\newblock in: \bibinfo{booktitle}{Proceedings of the 1st IEEE Symposium on
  Parallel and Distributed Processing}, pp. \bibinfo{pages}{202--209}.
\bibitem[{Keh and Lin(2000)}]{7}
\bibinfo{author}{H.-C. Keh}, \bibinfo{author}{J.-C. Lin},
\newblock \bibinfo{title}{On fault-tolerant embedding of hamiltonian cycles,
  linear arrays and rings in a flexible hypercube},
\newblock \bibinfo{journal}{Parallel Comput.} \bibinfo{volume}{26}
  (\bibinfo{year}{2000}) \bibinfo{pages}{769–781}.
\bibitem[{Youyao et~al.(2008)Youyao, Jungang, and Huimin}]{8}
\bibinfo{author}{L.~Youyao}, \bibinfo{author}{H.~Jungang},
  \bibinfo{author}{D.~Huimin},
\newblock \bibinfo{title}{A hypercube-based scalable interconnection network
  for massively parallel computing},
\newblock \bibinfo{journal}{Journal of Computers}  (\bibinfo{year}{2008}).
\bibitem[{Pai(2020)}]{24}
\bibinfo{author}{K.-j. Pai},
\newblock \bibinfo{title}{A parallel algorithm for constructing two
  edge-disjoint hamiltonian cycles in crossed cubes},
\newblock in: \bibinfo{booktitle}{International Conference on Algorithmic
  Applications in Management}, \bibinfo{organization}{Springer}, pp.
  \bibinfo{pages}{448--455}.
\bibitem[{Yang(2021)}]{34}
\bibinfo{author}{Y.~Yang},
\newblock \bibinfo{title}{Embedded connectivity of ternary n-cubes},
\newblock \bibinfo{journal}{Theoretical Computer Science}
  (\bibinfo{year}{2021}).
\bibitem[{El-Amawy and Latifi(1991)}]{2}
\bibinfo{author}{A.~El-Amawy}, \bibinfo{author}{S.~Latifi},
\newblock \bibinfo{title}{Properties and performance of folded hypercubes},
\newblock \bibinfo{journal}{IEEE Trans. Parallel Distrib. Syst.}
  \bibinfo{volume}{2} (\bibinfo{year}{1991}) \bibinfo{pages}{31–42}.
\bibitem[{Preparata and Vuillemin(1981)}]{3}
\bibinfo{author}{F.~P. Preparata}, \bibinfo{author}{J.~Vuillemin},
\newblock \bibinfo{title}{The cube-connected cycles: A versatile network for
  parallel computation},
\newblock \bibinfo{journal}{Commun. ACM} \bibinfo{volume}{24}
  (\bibinfo{year}{1981}) \bibinfo{pages}{300–309}.
\bibitem[{Efe(1992)}]{4}
\bibinfo{author}{K.~Efe},
\newblock \bibinfo{title}{The crossed cube architecture for parallel
  computation},
\newblock \bibinfo{journal}{IEEE Trans. Parallel Distrib. Syst.}
  \bibinfo{volume}{3} (\bibinfo{year}{1992}) \bibinfo{pages}{513–524}.
\bibitem[{Abuelrub(2008)}]{10}
\bibinfo{author}{E.~Abuelrub},
\newblock \bibinfo{title}{The hamiltonicity of crossed cubes in the presence of
  faults},
\newblock \bibinfo{journal}{Eng. Lett.} \bibinfo{volume}{16}
  (\bibinfo{year}{2008}) \bibinfo{pages}{453--459}.
\bibitem[{Chang et~al.(2000)Chang, Sung, and Hsu}]{11}
\bibinfo{author}{C.-P. Chang}, \bibinfo{author}{T.-Y. Sung},
  \bibinfo{author}{L.-H. Hsu},
\newblock \bibinfo{title}{Edge congestion and topological properties of crossed
  cubes},
\newblock \bibinfo{journal}{IEEE Transactions on Parallel and Distributed
  Systems} \bibinfo{volume}{11} (\bibinfo{year}{2000}) \bibinfo{pages}{64--80}.
\bibitem[{Fan et~al.(2005)Fan, Lin, and Jia}]{12}
\bibinfo{author}{J.~Fan}, \bibinfo{author}{X.~Lin}, \bibinfo{author}{X.~Jia},
\newblock \bibinfo{title}{Node-pancyclicity and edge-pancyclicity of crossed
  cubes},
\newblock \bibinfo{journal}{Information Processing Letters}
  \bibinfo{volume}{93} (\bibinfo{year}{2005}) \bibinfo{pages}{133--138}.
\bibitem[{Zhu et~al.(2007)Zhu, Xu, Hou, and Xu}]{13}
\bibinfo{author}{Q.~Zhu}, \bibinfo{author}{J.-M. Xu}, \bibinfo{author}{X.~Hou},
  \bibinfo{author}{M.~Xu},
\newblock \bibinfo{title}{On reliability of the folded hypercubes},
\newblock \bibinfo{journal}{Information Sciences} \bibinfo{volume}{177}
  (\bibinfo{year}{2007}) \bibinfo{pages}{1782--1788}.
\bibitem[{Kulasinghe and Bettayeb(1995)}]{22}
\bibinfo{author}{P.~Kulasinghe}, \bibinfo{author}{S.~Bettayeb},
\newblock \bibinfo{title}{Multiply-twisted hypercube with five or more
  dimensions is not vertex-transitive},
\newblock \bibinfo{journal}{Information Processing Letters}
  \bibinfo{volume}{53} (\bibinfo{year}{1995}) \bibinfo{pages}{33--36}.
\bibitem[{Wang et~al.(2021)Wang, Fan, Zhang, and Yu}]{35}
\bibinfo{author}{X.~Wang}, \bibinfo{author}{J.~Fan},
  \bibinfo{author}{S.~Zhang}, \bibinfo{author}{J.~Yu},
\newblock \bibinfo{title}{Node-to-set disjoint paths problem in cross-cubes},
\newblock \bibinfo{journal}{The Journal of Supercomputing}
  (\bibinfo{year}{2021}) \bibinfo{pages}{1--25}.
\bibitem[{Cheng et~al.(2017)Cheng, Wang, and Fan}]{25}
\bibinfo{author}{B.~Cheng}, \bibinfo{author}{D.~Wang},
  \bibinfo{author}{J.~Fan},
\newblock \bibinfo{title}{Constructing completely independent spanning trees in
  crossed cubes},
\newblock \bibinfo{journal}{Discrete Applied Mathematics} \bibinfo{volume}{219}
  (\bibinfo{year}{2017}) \bibinfo{pages}{100--109}.
\bibitem[{Pai et~al.(2020)Pai, Chang, Wu, and Chang}]{26}
\bibinfo{author}{K.-J. Pai}, \bibinfo{author}{R.-S. Chang},
  \bibinfo{author}{R.-Y. Wu}, \bibinfo{author}{J.-M. Chang},
\newblock \bibinfo{title}{Three completely independent spanning trees of
  crossed cubes with application to secure-protection routing},
\newblock \bibinfo{journal}{Information Sciences} \bibinfo{volume}{541}
  (\bibinfo{year}{2020}) \bibinfo{pages}{516--530}.
\bibitem[{Dong et~al.(2012)Dong, Zhou, Fu, and Yang}]{27}
\bibinfo{author}{Q.~Dong}, \bibinfo{author}{J.~Zhou}, \bibinfo{author}{Y.~Fu},
  \bibinfo{author}{X.~Yang},
\newblock \bibinfo{title}{Embedding a mesh of trees in the crossed cube},
\newblock \bibinfo{journal}{Information Processing Letters}
  \bibinfo{volume}{112} (\bibinfo{year}{2012}) \bibinfo{pages}{599--603}.
\bibitem[{Kulasinghe and Bettayeb(1995)}]{28}
\bibinfo{author}{P.~Kulasinghe}, \bibinfo{author}{S.~Bettayeb},
\newblock \bibinfo{title}{Embedding binary trees into crossed cubes},
\newblock \bibinfo{journal}{IEEE Transactions on Computers}
  \bibinfo{volume}{44} (\bibinfo{year}{1995}) \bibinfo{pages}{923--929}.
\bibitem[{Aschheim et~al.(2012)Aschheim, Femmam, and Zerarka}]{1}
\bibinfo{author}{R.~Aschheim}, \bibinfo{author}{S.~Femmam},
  \bibinfo{author}{M.~F. Zerarka},
\newblock \bibinfo{title}{New "graphiton" model: a computational discrete
  space, self-encoded as a trivalent graph},
\newblock \bibinfo{journal}{Comput. Inf. Sci.} \bibinfo{volume}{5}
  (\bibinfo{year}{2012}) \bibinfo{pages}{2--12}.
\bibitem[{Lin et~al.(2010)Lin, Yang, Hsu, and Chang}]{18}
\bibinfo{author}{J.-C. Lin}, \bibinfo{author}{J.-S. Yang},
  \bibinfo{author}{C.-C. Hsu}, \bibinfo{author}{J.-M. Chang},
\newblock \bibinfo{title}{Independent spanning trees vs. edge-disjoint spanning
  trees in locally twisted cubes},
\newblock \bibinfo{journal}{Information Processing Letters}
  \bibinfo{volume}{110} (\bibinfo{year}{2010}) \bibinfo{pages}{414--419}.
\bibitem[{Femmam and Zerarka(2017)}]{20}
\bibinfo{author}{S.~Femmam}, \bibinfo{author}{F.~M. Zerarka},
\newblock \bibinfo{title}{One-by-one embedding of the twisted hypercube into
  pancake graph},
\newblock in: \bibinfo{booktitle}{Building Wireless Sensor Networks},
  \bibinfo{publisher}{Elsevier}, \bibinfo{year}{2017}, pp.
  \bibinfo{pages}{145--169}.
\bibitem[{Matsubayashi(2015)}]{21}
\bibinfo{author}{A.~Matsubayashi},
\newblock \bibinfo{title}{Separator-based graph embedding into multidimensional
  grids with small edge-congestion},
\newblock \bibinfo{journal}{Discrete Applied Mathematics} \bibinfo{volume}{185}
  (\bibinfo{year}{2015}) \bibinfo{pages}{119--137}.

\end{thebibliography}






\vspace{0.2cm}
\noindent\textbf{Aymen Takie Eddine Selmi:} is a PhD student  in  University of Biskra, Algeria. He obtained  a master  degree in systems information, decision and optimization in 2018 from University of Biskra.His current research interests include parallel computing, optimization, and machine learning.
\vspace{0.2cm}

\noindent\textbf{Mohamed Faouzi Zerarka:} received Magister and PhD. Degrees in computer sciences department of computer sciences from university of Biskra, Algeria. He worked for a number of years in university of Batna Algeria, currently he is associate professor at Biskra university, Algeria, with expertise in parallel  architecture and computing.                                                           
\vspace{0.2cm}

\noindent\textbf{Abdelhakim Cheriet:} received the Engineering, M.Sc. and PhD degrees in 2004, 2008, and 2016 from the University of Biskra, Algeria, all in Computer Science. He worked as an assistant professor at the University of Biskra, Algeria until 2016. Currently, he is an associate professor at the University of Ouargla, Algeria. His research interests include: Optimization, Metaheuristics, and Evolutionary Computation.

\end{document}